\newif\iffigs\figstrue
\documentclass[a4paper,12pt]{article}
\usepackage{latexsym,amssymb,amsmath,lscape,graphics,setspace}
\usepackage{graphicx}        
\usepackage{longtable}
\usepackage{youngtab}
\usepackage{multirow}
\usepackage{color}
\usepackage{mathptmx}
\usepackage{slashed,epsfig}
\usepackage{amsfonts}

\newcommand{\e}{\textrm{e}}

\newcommand{\mathsym}[1]{{}}
\newcommand{\unicode}[1]{{}}
\newtheorem{definizione}{Definition}[section]

\newcommand{\bd}{\begin{definizione}}
\newcommand{\ed}{\end{definizione}}

\def\IC{\relax\,\hbox{$\inbar\kern-.3em{\rm C}$}}
\def\IG{\relax\,\hbox{$\inbar\kern-.3em{\rm G}$}}
\def\IB{\relax{\rm I\kern-.18em B}}
\def\ID{\relax{\rm I\kern-.18em D}}
\def\IL{\relax{\rm I\kern-.18em L}}
\def\IF{\relax{\rm I\kern-.18em F}}
\def\IH{\relax{\rm I\kern-.18em H}}
\def\II{\relax{\rm I\kern-.17em I}}
\def\IN{\relax{\rm I\kern-.18em N}}
\def\IP{\relax{\rm I\kern-.18em P}}
\def\IQ{\relax\,\hbox{$\inbar\kern-.3em{\rm Q}$}}
\def\bfzero{\relax\,\hbox{$\inbar\kern-.3em{\rm 0}$}}
\def\IK{\relax{\rm I\kern-.18em K}}
\def\IG{\relax\,\hbox{$\inbar\kern-.3em{\rm G}$}}
 \font\cmss=cmss10 \font\cmsss=cmss10 at 7pt
\def\IR{\relax{\rm I\kern-.18em R}}
\def\ZZ{\relax\ifmmode\mathchoice
{\hbox{\cmss Z\kern-.4em Z}}{\hbox{\cmss Z\kern-.4em Z}} {\lower.9pt\hbox{\cmsss Z\kern-.4em Z}}
{\lower1.2pt\hbox{\cmsss Z\kern-.4em Z}}\else{\cmss Z\kern-.4em Z}\fi}
\def\bfone{\relax{\rm 1\kern-.35em 1}}

\def\n010{N^{0,1,0}}
\def\inbar{\vrule height1.5ex width.4pt depth0pt}
\def\bfzero{\relax{\rm I\kern-.18em 0}}
\def\bfone{\relax{\rm 1\kern-.35em 1}}

\DeclareFontFamily{U}{rsf}{} \DeclareFontShape{U}{rsf}{m}{n}{
  <5> <6> rsfs5 <7> <8> <9> rsfs7 <10-> rsfs10}{}
\DeclareMathAlphabet\Scr{U}{rsf}{m}{n}

\def\e{\epsilon} 
 \def\l{\lambda}

\newcommand{\ft}[2]{{\textstyle\frac{#1}{#2}}}

\def\l{\lambda}
\def\1bar{1\hskip -.275cm -}
\def\2bar{2\hskip -.275cm -}
\def\3bar{3\hskip -.275cm -}

\newsavebox{\uuunit}
\sbox{\uuunit}
                 {\setlength{\unitlength}{0.825em}
                      \begin{picture}(0.6,0.7)
                                      \thinlines
                                      \put(0,0){\line(1,0){0.5}}
                                      \put(0.15,0){\line(0,1){0.7}}
                                      \put(0.35,0){\line(0,1){0.8}}
                                     \multiput(0.3,0.8)(-0.04,-0.02){10}{\rule{0.5pt}{0.5pt}}
                      \end {picture}}

\makeatletter \@addtoreset{equation}{section} \makeatother

\def\bfone{\relax{\rm 1\kern-.35em 1}}

\def\bfone{\relax{\rm 1\kern-.35em 1}}
\font\cmss=cmss10 \font\cmsss=cmss10 at 7pt

\newcommand{\so}{\mathfrak{so}}
\newcommand{\su}{\mathfrak{su}}

\def\bfone{\relax{\rm 1\kern-.35em 1}}
\def\inbar{\vrule height1.5ex width.4pt depth0pt}
\def\IC{\relax\,\hbox{$\inbar\kern-.3em{\rm C}$}}
\def\ID{\relax{\rm I\kern-.18em D}}
\def\IF{\relax{\rm I\kern-.18em F}}
\def\IH{\relax{\rm I\kern-.18em H}}
\def\II{\relax{\rm I\kern-.17em I}}
\def\IN{\relax{\rm I\kern-.18em N}}
\def\IP{\relax{\rm I\kern-.18em P}}
\def\IQ{\relax\,\hbox{$\inbar\kern-.3em{\rm Q}$}}
\def\IR{\relax{\rm I\kern-.18em R}}
\font\cmss=cmss10 \font\cmsss=cmss10 at 7pt
\def\ZZ{\relax\ifmmode\mathchoice
{\hbox{\cmss Z\kern-.4em Z}}{\hbox{\cmss Z\kern-.4em Z}} {\lower.9pt\hbox{\cmsss Z\kern-.4em Z}}
{\lower1.2pt\hbox{\cmsss Z\kern-.4em Z}}\else{\cmss Z\kern-.4em Z}\fi}

\def\e{\epsilon} 
 \def\l{\lambda}

\def\bar{\overline}

\def\hat{\widehat}

\def\Coe#1.#2.{{#1\over #2}}

\def\coe#1.#2.{\relax{\textstyle {#1 \over #2}}\displaystyle}

\def\to{\rightarrow}
\def\notin{\hbox{{$\in$}\kern-.51em\hbox{/}}}

\def\IE{\relax{{\rm I\kern-.18em E}}}

\def\IGam{\relax{{\rm I}\kern-.18em \Gamma}}

\def\IA{\relax{\hbox{{\rm A}\kern-.82em {\rm A}}}}

\begin{document}
\begin{titlepage}
\begin{center}
\begin{flushright}
ARC-17-01\\
YITP-17-48
\end{flushright}
\vskip 0.2cm
\vskip 0.2cm
{\Large  \bf The Integral Form of D=3 Chern-Simons Theories \\ \vskip .5cm
Probing ${\mathbb C}^n/\Gamma$ Singularities}\\[1cm]
 {\Large P.~Fr\'e}${}^{\; a,b,c,d,}$\footnote{Prof. Fr\'e is presently fulfilling the duties of Scientific Counselor of
the Italian Embassy in the Russian Federation, Denezhnij pereulok, 5, 121002 Moscow, Russia.}
\footnote{pietro.fre@esteri.it}
{\Large and  P.A. Grassi}${}^{\;b,c,e,f,}$\footnote{pietro.grassi@uniupo.it}
\\[10pt]
{${}^a$\sl\small Dipartimento di Fisica, Universit\`a di Torino, via P. Giuria 1, \ 10125 Torino \ Italy\\}
\vspace{5pt}
{${}^b$\sl\small INFN -- Sezione di Torino, via P. Giuria 1, \ 10125 Torino \ Italy\\}
\vspace{5pt}
{${}^c$\sl\small Arnold-Regge Center, via P. Giuria 1, \ 10125 Torino \ Italy\\}
\vspace{5pt}
{{\em $^d$\sl\small  National Research Nuclear University MEPhI, (Moscow Engineering Physics Institute),}}\\
{\sl Kashirskoye shosse 31, 115409 Moscow, Russia} \\
\vspace{5pt}
{${}^e$\sl\small DISIT, Universit\`a del Piemonte Orientale,
via T. Michel, 11, \ 15121 Alessandria \ Italy\\}
\vspace{5pt}
{${}^f$\sl\small
Center for Gravitational Physics, Yukawa Institute for Theoretical Physics, \\
Kyoto University,
Kyoto 606-8502, Japan}

~\quad\\
\quad \\
\quad \vspace{6pt}
\vspace{15pt}
\begin{abstract}
{We consider D=3 supersymmetric Chern Simons gauge theories both from the point of view of their formal
structure and of their applications to the $\mathrm{AdS_4/CFT_3}$ correspondence. From the structural
view-point, we use the new formalism of integral forms in superspace that utilizes the rheonomic Lagrangians
and the Picture Changing Operators, as an algorithmic tool providing the connection between different
approaches to supersymmetric theories. We provide here the generalization to an arbitrary K\"ahler manifold
with arbitrary gauge group and arbitrary superpotential of the rheonomic lagrangian of D=3 matter coupled
gauge theories constructed years ago. From the point of view of the $\mathrm{AdS_4/CFT_3}$ correspondence and
more generally of M2-branes we emphasize the role of the K\"ahler quotient data in determining the field
content and the interactions of the Cherns Simons gauge theory when the transverse space to the brane is a
non-compact K\"ahler quotient $K_4$ of some flat variety with respect to a suitable group. 
The crepant resolutions of ${\mathbb C}^n/\Gamma$ singularities fall in this category. In the present paper we anticipate the
general scheme how the geometrical data are to be utilized in the construction of the D=3 Chern-Simons Theory
supposedly dual to the corresponding M2-brane solution.}
\end{abstract}
\end{center}
\end{titlepage}
\tableofcontents \noindent {}
\newpage
\def\nslash{\nabla\!\!\!\!/}
\section{Conceptual and Historical Introduction}
The vision of the AdS/CFT correspondence has its starting point in November 1997 with the publication on the
ArXive of  a paper by Juan Maldacena \cite{Maldacena:1997re} on the large $N$ limit of gauge theories.
\par
From the viewpoint of the superstring scientific community this was seen as the first explicit example of the
long sought  duality between gauge theories and superstrings. Yet the scope of this correspondence was
destined to be enlarged in many directions and to become, more generically, the \textit{gauge/gravity
correspondence} based on various declinations of the basic idea referred to as the \textit{holographic
principle}. According to this latter, fundamental informations on the quantum behavior of fields leaving on
some boundary of a larger space-time can be obtained from the classical gravitational dynamics of fields
leaving in the bulk of that space-time. Such wider approach to the AdS/CFT correspondence
diminishes the emphasis on strings and brings to higher
relevance both supergravity theories and their perturbative and non-perturbative symmetries. In such a
framework geometrical issues become the central focus of attention.
\par
It followed immediately, from december 1997 to the late spring of 1998, a series of fundamental papers by
Ferrara, Fronsdal, Zaffaroni, Kallosh and Van Proeyen
\cite{Ferrara:1997dh},\cite{Ferrara:1998ej},\cite{Kallosh:1998ph},\cite{Ferrara:1998jm}, where the algebraic
and field theoretical basis of the correspondence was clarified independently from microscopic string
considerations.
\par
The AdS/CFT correspondence has a relative simple origin which, however,
is extremely rich in ramified and powerful consequences. The key point is the double interpretation of any
anti de Sitter group $\mathrm{SO(2,p+1)}$ as the isometry group of the $\mathrm{AdS_{p+2}}$ space or as the
conformal group on the $(p+1)$-dimensional boundary $\partial\mathrm{AdS_{p+2}}$. Such a double
interpretation is inherited by the supersymmetric extensions of $\mathrm{SO(2,p+1)}$. This is what leads to
consider \textit{superconformal field theories} on the boundary. Two cases are of particular relevance
because of concurrent reasons which are peculiar to them: from the algebraic side the essential use of one of
the low rank sporadic isomorphisms of orthogonal Lie algebras, from the supergravity side the existence of a
spontaneous compactification of the Freund-Rubin type \cite{freurub}. The two cases are:
\begin{description}
  \item[A)] The case $p=3$ which leads to $\mathrm{AdS_5}$ and to its $4$-dimensional boundary. Here
  the sporadic isomorphism  is $\mathrm{SO(2,4)}\sim \mathrm{SU(2,2)}$ which implies that
   the  list of superconformal algebras is given by the superalgebras $\su(2,2 \mid \mathcal{N})$
  for $1\leq \mathcal{N} \leq 4$. On the other hand in Type IIB Supergravity, there is a self-dual five-form
  field strength. Giving a v.e.v to this latter ($F_{a_1a_2a_3a_4a_5} \ltimes \epsilon_{a_1a_2a_3a_4a_5}$),
  one splits the ambient ten-dimensional space into $5\oplus5$ where the first $5$ stands for the $\mathrm{AdS_5}$ space,
  while the second $5$ stands for any compact $5$-dimensional Einstein manifold $\mathcal{M}_5$.
  The holonomy of the metric cone on the latter $\mathcal{C}(\mathcal{M}_5)$ decides
  the number of supersymmetries and on the $4$-dimensional boundary
   $\partial \mathrm{AdS_5}$ we have a superconformal Yang-Mills gauge theory.
  \item[B)] The case $p=2$ which leads  to $\mathrm{AdS_4}$ and to its $3$-dimensional boundary.
  Here the sporadic isomorphism  is $\mathrm{SO(2,3)}\sim \mathrm{Sp(4,\mathbb{R})}$ which implies that
  the  list of superconformal algebras is given by the superalgebras $\mathrm{Osp}(\mathcal{N} \mid 4)$
  for $\mathcal{N} = 1, 2, 3, 6, 8$. On the other hand in $D=11$ Supergravity, there is a a four-form
  field strength. Giving a v.e.v to this latter ($F_{a_1a_2a_3a_4} \ltimes \epsilon_{a_1a_2a_3a_4}$),
  one splits the ambient ten-dimensional space into $4\oplus7$ where  $4$ stands for the $\mathrm{AdS_4}$ space,
  while  $7$ stands for any compact $7$-dimensional Einstein manifold $\mathcal{M}_7$.
  The holonomy of the metric cone on the latter $\mathcal{C}(\mathcal{M}_7)$
  decides the number of supersymmetries and on the $3$-dimensional boundary
  $\partial \mathrm{AdS_4}$ we should have a superconformal  gauge theory.
\end{description}
The first case was that mostly explored at the beginning of the
AdS/CFT correspondence in 1998 and in successive years. Yet the existence of
the second case was immediately evident to anyone who had experience
in supergravity and particularly to those who had worked in
Kaluza-Klein supergravity in the years 1982-1985. Thus in a series
of papers
\cite{Fabbri:1999mk},\cite{sergiotorino},\cite{Fabbri:1999ay},\cite{Fre':1999xp},\cite{Fabbri:1999hw},\cite{ringoni},\cite{Billo:2000zs},
mostly produced by the Torino Group and by the SISSA Group, the $\mathrm{AdS_4/CFT_3}$
correspondence was proposed and intensively developed in the spring
and in the summer of the year 1999. One leading idea, motivating
this outburst of activity, was that the entire corpus of results on
Kaluza-Klein mass-spectra which had been derived in the years
1982-1986, \cite{freurub},
\cite{round7a},\cite{osp48},\cite{squash7a},\cite{biran},\cite{gunawar},\cite{kkwitten},
\cite{noi321},\cite{spec321},\cite{multanna},\cite{pagepopeM},\cite{dafrepvn},\cite{pagepopeQ},
\cite{univer},\cite{bosmass},\cite{frenico},\cite{castromwar}, could
now be recycled in the new superconformal interpretation. Actually
it was immediately clear that the Kaluza-Klein towers of states, in
particularly those corresponding to short representations of the
superalgebra $\mathrm{Osp}(\mathcal{N}\mid 4)$, provided an
excellent testing ground for the $\mathrm{AdS_4/CFT_3}$
correspondence. One had to conceive candidate superconformal field
theories living on the boundary, that were able to reproduce all the
infinite towers of Kaluza Klein multiplets  as corresponding towers
of composite operators with the same quantum numbers.
\par
In the case the manifold $\mathcal{M}_7$ was a coset manifold
$\mathcal{G/H}$, an exhaustive list of cases was known since the
middle eighties, thanks to the work of Castellani, Romans and Warner
\cite{castromwar}. The supersymmetric cases form an even shorter
sublist of the main list in \cite{castromwar} and were also
classified by the same authors (see table \ref{sasakiani} and table
\ref{n=1casi}).
\par
Since it was clear that the theory on the boundary had to be a
matter coupled gauge-theory, in three papers \cite{Fabbri:1999ay},
\cite{ringoni} and \cite{Fre':1999xp}, the general form of  matter
coupled $\mathcal{N}=2,3$ non abelian gauge theories in D=3, with
both a canonical kinetic term for the gauge fields and a Chern
Simons one, were constructed using auxiliary fields and the
rheonomic approach.
\par
In the series of papers
\cite{Fabbri:1999mk},\cite{Fabbri:1999ay},\cite{Fre':1999xp},\cite{Fabbri:1999hw},\cite{ringoni},\cite{Billo:2000zs},
it was also conjectured that the gauge theories \textit{dual to the
supergravity backgrounds} of type $\mathrm{AdS_4 }\times
\mathcal{M}_7$ have an infrared fixed point where the Yang Mills
coupling constant goes to infinity.  In this limit the kinetic terms
are removed for  all the fields  in the  gauge multiplet. These
latter  become auxiliary fields and, with the exception of the non
abelian gauge one-forms, they can be integrated away leaving, as
remnant, a pure Chern Simons gauge theory with a very specific form,
that was discussed in the quoted  papers.
\par
The question remains how to fill the black box of matter multiplets
in the general Chern Simons lagrangian constructed in the way
sketched above. We address this issue in the  subsection after the
next, yet before doing that we clarify the general scope of the
present paper.
\subsection{The scope and the goals of the present paper}
In view of the above considerations, the scope of the present paper is an in depth analysis of matter coupled
Maxwell Chern-Simons supersymmetric gauge theories in three space-time dimensions. We aim at a general scheme
that encompasses also $\mathcal{N}=3$ and $\mathcal{N}=6$ Chern-Simons theories that are the basis of the
ABJM-model.
\par
From the point of view of the contents of the theory we are particularly interested in candidates for the
dual CFT.s of M2-brane solutions of D=11 supergravity probing $\mathbb{C}^n/\Gamma$ singularities and their
resolutions. From the point of view of the constructive principles of supersymmetric field-theories, we are
particulary interested in the recently discovered set-up of integral forms in superspace
\cite{Castellani:2015paa,Catenacci:2016qzd,Grassi:2016apf,Castellani:2016ibp} and we plan to explore the
properties of the considered class of gauge theories in this respect. We will first provide the appropriate
generalization of the rheonomic lagrangian derived in \cite{Fabbri:1999ay} to an arbitrary K\"ahler manifold
with an arbitrary triholomorphic isometry group. Then, according with the general views introduced in
\cite{Castellani:2015paa}, we plan to show that the space--time lagrangian and other superfield formulations of
the same theory can be obtained by multiplying the rheonomic lagrangian with suitable closed integral forms
belonging to the same cohomology class and by integrating on full superspace the result of this wedge
multiplication.
\par
The class of considered gauge theories is particularly suited to
explore the new view point on superspace since they have a finite
set of auxiliary fields and the rheonomic action is an off-shell
closed form.
\subsection{The Sasakian structure and the metric cone}
Coming back to the question \textit{how to fill the black box of
matter multiplets in the general Chern Simons lagrangian} we note
that it is in the resolution of this problem that the interplay
between the geometry of the compactification manifold
$\mathcal{M}_7$ and the structure of the $d=3$ superconformal field
theory becomes evident.
\par
In paper \cite{Fabbri:1999hw} the authors introduced a systematic bridge between the geometry of
$\mathcal{M}_7$ and the structure of the boundary gauge theory based on the crucial observation that all the
$7$-dimensional cosets with at least two Killing spinors of the
$\mathrm{AdS}$-type are sasakian manifolds or
tri-sasakian manifolds.
\par
What sasakian means is visually summarized in the following table.
\begin{center}
\begin{tabular}{|cccc|}
  \hline
 base  of the fibration &   projection & $7$-manifold &  metric cone  \\
  $\mathcal{B}_6$ & $\stackrel{\pi}{\longleftarrow}$ &$\mathcal{M}_7$ & $\mathcal{C}\left( \mathcal{M}_7\right)$ \\
  $\Updownarrow$ & $\forall p \in \mathcal{B}_6 \quad \pi^{-1}(p) \, \sim \, \mathbb{S}^1 \,$
   &$\Updownarrow$ & $\Updownarrow$ \\
  K\"ahler $K_3$ &$\null$ & sasakian& K\"ahler Ricci flat $K_4$ \\
  \hline
\end{tabular}
\end{center}
First of all the $\mathcal{M}_7$ manifold must admit an
$\mathbb{S}^1$-fibration over a complex K\"ahler three-fold $K_3$:
\begin{equation}\label{fibratoS1}
    \pi \quad : \quad \mathcal{M}_7 \stackrel{\mathbb{S}^1}{\longrightarrow} \, K_3
\end{equation}
Calling $z^i$ the three complex coordinates of $K_3$ and $\phi$ the
angle spanning $\mathbb{S}^1$, the fibration  means that the metric
of $\mathcal{M}_7$ admits the following representation:
\begin{equation}\label{fibrametricu}
    ds^2_{\mathcal{M}_7}\, = \, \left(d\phi - \mathcal{A}\right)^2 \, + \, g_{ij^\star} \, dz^i \otimes d{\bar z}^{j^\star}
\end{equation}
where the one--form $\mathcal{A}$ is some suitable connection
one--form on the $\mathrm{U(1)}$-bundle (\ref{fibratoS1}).
\par
Secondly the metric cone $\mathcal{C}\left( \mathcal{M}_7\right)$
over the manifold $\mathcal{M}_7$ defined by the direct product
$\mathbb{R}_+\otimes \mathcal{M}_7$ equipped with the following
metric :
\begin{equation}\label{gustoconetto}
    ds^2_{\mathcal{C}\left( \mathcal{M}_7\right)} \, =\,dr^2 + 4 \,e^2 \, r^2 \, ds^2_{\mathcal{M}_7}
\end{equation}
should also be a Ricci-flat complex K\"ahler $4$-fold. In the above
equation $e$ just denotes a constant scale parameter with the
dimensions of an inverse length $\left[e\right] = \ell^{-1}$.
\par
Altogether the Ricci flat Kahler manifold $K_4$, which plays the
role of transverse space to the M2-branes, is a line-bundle over the
base manifold $K_3$:
\begin{eqnarray}\label{convecchio}
    \pi &\quad : \quad& K_4 \, \longrightarrow \, K_3 \nonumber\\
     \forall p \in K_3 && \pi^{-1}(p) \, \sim \, \mathbb{C}^\star
\end{eqnarray}
 All the manifolds listed in table \ref{sasakiani} are sasakian in the sense described above. The
$\so(8)$-holonomy mentioned in this table is the holonomy of the
Levi-Civita connection of the metric cone $\mathcal{C}\left
(\mathcal{M}_7\right)$ which can be easily calculated from that of
the $\mathcal{M}_7$-manifold relying on the following one-line
construction. Define the vielbein of $\mathcal{C}\left(
\mathcal{M}_7\right)$ in terms of the vielbein of $\mathcal{M}_7$ in
the following way:
\begin{equation}\label{vinileto}
    V^I \, = \, \left\{\begin{array}{rcl}
                         V^0 & = & dr \\
                         V^\alpha & =  & e \, r \, \mathcal{B}^\alpha
                       \end{array}
     \right. \quad r \in \mathbb{R}_+
\end{equation}
where $ds^2_{\mathcal{M}_7} \, = \, \sum_{\alpha =1}^7
\mathcal{B}^\alpha \otimes \mathcal{B}^\alpha$. The torsion
equation:
\begin{equation}\label{connettus}
    dV^I \, + \, \Omega^{IJ} \wedge V^J \, = \, 0
\end{equation}
where $\Omega^{IJ}$ is the spin--connection of the metric cone, is
solved by:
\begin{eqnarray}\label{corinna}
    \Omega^{\alpha\beta} & = & \mathcal{B}^{\alpha\beta}\nonumber\\
 \Omega^{0\beta} & = & - 2\, e \, r \, \mathcal{B}^{\beta}
\end{eqnarray}
having denoted by $\mathcal{B}^{\alpha\beta}$ the spin--connection
of $\mathcal{M}_7$, namely $d\mathcal{B}^{\alpha} \, +
\,\mathcal{B}^{\alpha\beta}\wedge \mathcal{B}^{\beta} \, = \, 0$.
According to the summary of Kaluza--Klein supergravity presented in
\cite{antonpietroads}, $\Omega^{IJ}$ is the $\so(8)$-connection
whose holonomy decides the number of Killing spinor admitted by the
$\mathrm{AdS_4} \times \mathcal{M}_7$ compactification of M-theory.
When this holonomy vanishes we have the maximal number of preserved
supersymmetries. When it is $\mathrm{SU(3)}\subset \mathrm{SO(8)}$
we have $\mathcal{N}=2$. When it is $\mathrm{SU(2)}\subset
\mathrm{SO(8)}$ we might in principle expect $\mathcal{N}=4$, but we
actually have only $\mathcal{N}=3$, as firstly remarked by
Castellani, Romans and Warner in 1985.
\par
In \cite{Fabbri:1999hw},  it was emphasized that the fundamental
geometrical clue to the field content of the \textit{superconformal
gauge theory} on the boundary is provided by the construction of the
K\"ahler manifold $K_4$ as a holomorphic algebraic variety in some
higher dimensional affine or projective space $\mathbb{V}_{q}$, plus
a K\"ahler quotient. The equations identifying the algebraic locus
in $\mathbb{V}_{q}$ are related with the superpotential $W$
appearing in the $d=3$ lagrangian, while the K\"ahler quotient is
related with the $D$-terms appearing in the same lagrangian. The
coordinates $u,v$ of the space $\mathbb{V}_{q}$ are the scalar
fields of the \textit{superconformal gauge theory}, whose vacua,
namely the set of extrema of its scalar potential, should be in
one-to-one correspondence with the points of $K_4$. Going from one
to multiple M2--branes just means that the coordinate $u,v$ of
$\mathbb{V}_{q}$ acquire color indices under a proper set of color
gauge groups and are turned into matrices. In this way we obtain
\textit{quivers}.
\begin{table}
  \centering
  {\small \begin{tabular}{|c||c|c|c|l|}
\hline
 $\mathcal{N}$ & Name & Coset &$\begin{array}{c}
   \mbox{Holon.} \\
   \so(8) \mbox{ bundle } \
 \end{array}$ & Fibration \\
  \hline
  8 & $\mathbb{S}^7$ & $\frac{\mathrm{SO(8)}}{\mathrm{SO(7)}}$ & 1 &
  $ \left \{ \begin{array}{l}
    \mathbb{S}^7 \, \stackrel{\pi}{\Longrightarrow} \, \mathbb{P}^3 \\
    \forall \, p \, \in \, \mathbb{P}^3 \, ; \, \pi^{-1}(p) \, \sim \, \mathbb{S}^1\\
  \end{array}  \right. $ \\
  \hline 2 & $M^{1,1,1}$ & $\frac{\mathrm{SU(3)\times SU(2)\times U(1)}}{\mathrm{SU(2) \times U(1) \times U(1)
  }}$ & $\mathrm{SU(3)}$ & $ \left \{ \begin{array}{l}
    M^{1,1,1} \, \stackrel{\pi}{\Longrightarrow} \, \mathbb{P}^2  \, \times \, \mathbb{P}^1\\
    \forall \, p \, \in \, \mathbb{P}^2  \, \times \, \mathbb{P}^1\, ; \, \pi^{-1}(p) \, \sim \, \mathbb{S}^1\\
  \end{array}  \right. $ \\
  \hline
   2 & $Q^{1,1,1}$ & $\frac{\mathrm{SU(2)\times SU(2)\times SU(2) \times U(1)}}{\mathrm{U(1) \times U(1) \times
  U(1) }}$ & $\mathrm{SU(3)}$ & $ \left \{ \begin{array}{l}
    Q^{1,1,1} \, \stackrel{\pi}{\Longrightarrow} \, \mathbb{P}^1  \, \times \, \mathbb{P}^1\, \times \,\mathbb{P}^1  \\
    \forall \, p \, \in \, \mathbb{P}^1  \, \times \, \mathbb{P}^1\, \times \,\mathbb{P}^1  \, ; \,
    \pi^{-1}(p) \, \sim \, \mathbb{S}^1\\
  \end{array}  \right. $ \\
  \hline
   2 & $V^{5,2}$ & $\frac{\mathrm{SO(5)}}{\mathrm{SO(2)}}$ & $\mathrm{SU(3)}$ &
  $ \left \{ \begin{array}{l}
    V^{5,2} \, \stackrel{\pi}{\Longrightarrow} \, M_a \, \sim \, \mbox{quadric in } \mathbb{P}^4  \\
    \forall \, p \, \in \, \, M_a \, \, ; \, \pi^{-1}(p) \, \sim \, \mathbb{S}^1\\
  \end{array}  \right. $ \\
  \hline
   3 & $N^{0,1,0}$ & $\frac{\mathrm{SU(3)\times SU(2)}}{\mathrm{SU(2)\times U(1)}}$ & $\mathrm{SU(2)}$ &
  $ \left \{ \begin{array}{l}
    N^{0,1,0} \, \stackrel{\pi}{\Longrightarrow} \, \mathbb{P}^2  \\
    \forall \, p \, \in \, \, \mathbb{P}^2 \, \, ; \, \pi^{-1}(p) \, \sim \, \mathbb{S}^3\\
    \hline
    N^{0,1,0} \, \stackrel{\pi}{\Longrightarrow} \, \frac{\mathrm{SU(3)}}{\mathrm{U(1)}\times \mathrm{U(1)}}  \\
    \forall \, p \, \in \, \, \frac{\mathrm{SU(3)}}{\mathrm{U(1)}\times \mathrm{U(1)}} \, \, ; \, \pi^{-1}(p)
    \, \sim \, \mathbb{S}^1
  \end{array}  \right. $ \\
  \hline
  \end{tabular}}
  \caption{The homogeneous $7$-manifolds that admit at least $2$ Killing spinors are all sasakian or
  tri-sasakian. This is evident from the fibration structure of the $7$-manifold, which is either a fibration
  in circles $\mathbb{S}^1$ for the $\mathcal{N}=2$ cases or a fibration in $\mathbb{S}^3$ for the unique
  $\mathcal{N}=3$ case corresponding to the $\mathrm{N}^{0,1,0}$ manifold. Since this latter is also an $\mathcal{N}=2$
  manifold, there is in addition the $\mathbb{S}^1$ fibration.}\label{sasakiani}
\end{table}
\begin{table}
  \centering
  {\small \begin{tabular}{|c||c|c|c|}
\hline
 $\mathcal{N}$ & Name & Coset &$\begin{array}{c}
   \mbox{Holon.} \\
   \so(8) \mbox{ bundle } \
 \end{array}$ \\
  \hline
   \null & \null & \null & \null \\
  1 & $\mathbb{S}^7_{squashed}$ & $\frac{\mathrm{SO(5)\times SO(3)}}{\mathrm{SO(3)\times SO(3)}}$ & $\mathrm{SO(7)}^+$  \\
  \null & \null & \null & \null \\
  \hline
   \null & \null & \null & \null \\
  1 & $\mathrm{N^{p,q,r}}$ & $\frac{\mathrm{SU(3))\times U(1)}}{\mathrm{U(1) \times U(1) }}$ & $\mathrm{SO(7)}^+$  \\
   \null & \null & \null & \null \\
  \hline
  \end{tabular}}
  \caption{The homogeneous $7$-manifolds that admit just one Killing spinors are the squashed $7$-sphere and the infinite
  family of $\mathrm{N^{p,q,r}}$ manifolds for $p,q,r \ne 0,1,0$. }\label{n=1casi}
\end{table}
\par
All these conceptual and algorithmic points were enumerated in the
set of papers
\cite{Fabbri:1999mk},\cite{Fabbri:1999ay},\cite{Fre':1999xp},
\cite{Fabbri:1999hw},\cite{Billo:2000zs}, where the cases
$Q^{1,1,1}$, $M^{1,1,1}$ and $N^{0,1,0}$ were worked out in detail,
finding the algebraic embedding, defining the superpotential and the
quiver. Finally the Kaluza--Klein spectrum of supergravity
compactified on each of these three spaces was matched with the
spectrum of composite conformal operators in the corresponding
boundary superconformal theory.
\subsection{Resurrection of the $AdS_4/CFT_3$ correspondence and the ABJM setup}
The subject of the $\mathrm{AdS_3\times CFT_3}$  correspondence was resurrected ten years later in 2007-2009
by the work presented in papers \cite{Gaiotto:2007qi},\cite{Aharony:2008ug},\cite{Bagger:2007vi} which
stirred a great interest in the scientific community and obtained a very large number of citations. We
confess that the formalism of three-algebras introduced in \cite{Bagger:2007vi} is not very clear to us, but
we rely on the statement by the authors of \cite{Aharony:2008ug} that their construction is completely
equivalent to the theory presented in \cite{Bagger:2007vi}. The ABJM-construction of \cite{Aharony:2008ug} is
instead very clear and the attentive reader, making the required changes of notations and names of the
objects, can verify that the $\mathcal{N}=3$ lagrangian presented there is just the same as that constructed
in papers \cite{Fabbri:1999mk},\cite{Fabbri:1999ay} by letting the Yang-Mills coupling constant go to
infinity. What is new and extremely important in the ABJM model is the relative quantization of the Chern
Simons levels $k_{1,2}$ of the two gauge groups and its link to a quotiening of the seven sphere by means of
a cyclic group $\mathbb{Z}_k$. Indeed the theories presented in \cite{Aharony:2008ug} pertain to the first
case in table \ref{sasakiani}, modified by a \textit{finite group quotiening}. We just regret that the
authors of \cite{Aharony:2008ug} did not consider it appropriate to quote the papers of ten years before that
contain a large part of the ground basis of their results.
\par
For this reason in the first sections of these notes we review the
constructions of \cite{Fabbri:1999mk},\cite{Fabbri:1999ay},
translated into a more modern notation that refers from the
beginning to those geometrical structures, K\"ahler metrics and
moment maps, we will utilize in the sequel. By these means we want
to show that the  construction method introduced in
\cite{Aharony:2008ug} is just identical to that laid down in
1998-1999 in the several times quoted series of papers; furthermore
we  want to fix the framework where to discuss and possibly answer a
new question which we presently formulate.
\subsection{Finite group quotiening}
As we emphasized the key guiding item in the construction of the d=3
gauge theory is the $\mathrm{K_4}$ manifold and its representation
as an algebraic locus in some $\mathbb{V}_q$. We can extract the
logic which underlies \cite{Aharony:2008ug}, by means of the
following arguments. First consider the following  projections and
embeddings pertaining to the case where $\mathcal{M}_7$ is a smooth
coset manifold
\begin{equation}\label{caluffo}
  K_3 \, \stackrel{\pi}{\longleftarrow} \, \mathcal{G/H} \,  \stackrel{C}{ \hookrightarrow} \,  K_4 \,
  \stackrel{A}{ \hookrightarrow} \, \mathbb{V}_q
\end{equation}
In the above formula $\stackrel{C}{ \hookrightarrow}$ is the
embedding map into the metric cone, while $\stackrel{A}{
\hookrightarrow}$ denotes the algebraic embedding into an affine or
projective variety by means of a suitable set of algebraic
equations.
\par
For instance in the case of the seven sphere $\mathcal{G/H}\, = \,
\mathrm{SO(8)/SO(7)}$, we have $K_3 \, = \, \mathbb{P}^3$ and $K_4
\, = \, \mathbb{C}^4 \sim \mathbb{R}^8$. Then the algebraic map
$\stackrel{A}{ \hookrightarrow}$ is just the identity map since
$\mathbb{V}_q \, = \, \mathbb{C}^4$.
\par
On the contrary, in the case  $N^{0,1,0}$, the base manifold $K_3 \, =
\frac{\mathrm{SU(3)}}{\mathrm{U(1)\times U(1)}}$ is just the $\su(3)$ flag manifold and $K_4$ is obtained as
the K\"ahler quotient of an algebraic locus cut in $\mathbb{V}_q \, = \, \mathbb{C}^6$ by a quadric equation.
In this particular case the entire procedure how to go from $\mathbb{C}^6$ to $K_4$ can be seen as a
HyperK\"ahler quotient with respect to the triholomorphic action of a $\mathrm{U(1)}$ group:
\begin{equation}\label{ipercalloquozio}
  K_4 \, = \, \mathbb{C}^6 //_H \, \mathrm{U(1)}
\end{equation}
The quadric constraint is traced back to the vanishing of the
holomorphic part of the triholomorphic moment map, while the
K\"ahler quotient encodes the constraint coming from the real part
of the same moment map.
\par
Next we consider some finite group $\Gamma \subset \mathcal{G}$ and
in eq.(\ref{caluffo}) we replace  the homogeneous space
$\mathcal{G/H}$ with the orbifold $\frac{\mathcal{G/H}}{\Gamma}$.
The finite group quotient extends both to the projection map and to
the metric cone enlargment. Thus eq.(\ref{caluffo}) is replaced by:
\begin{equation}\label{sberleffo}
  \frac{K_3}{\Gamma} \, \stackrel{\pi}{\longleftarrow} \, \frac{\mathcal{G/H}}{\Gamma} \,
  \stackrel{C}{ \hookrightarrow} \,  \frac{K_4}{\Gamma} \,
  \stackrel{A}{ \hookrightarrow} \, \mathbb{V}_q
  \end{equation}
Typically the quotient $\frac{K_4}{\Gamma}$ is a singular manifold.
We need a resolution of the singularities by means of an appropriate
resolving map:
\begin{equation}\label{cubertollo}
  X^{\mathrm{res}} \, \rightarrow \,\frac{ K_4}{\Gamma}
\end{equation}
which typically leads to an affine variety $X^{\mathrm{res}} \,
\stackrel{A}{\hookrightarrow} \, \mathbb{C}^q$ embedded by suitable
algebraic equations into some $\mathbb{C}^q$.
\par
The final outcome is that the coordinates $z^i$ of $\mathbb{C}^q$ are the matter fields in the $d=3$
conformal field theory, while the embedding equations should determine the superpotential $\mathcal{W}(z)$.
The gauging is instead dictated by the final K\"ahler quotient of the resolved algebraic variety
$X^{\mathrm{res}}$ which produces the resolved metric cone $K_4^{\mathrm{res}}$.
\subsection{Crepant resolution of Gorenstein singularities}
It appears from the above discussion that the most fundamental
question at stake is a classical problem of algebraic geometry,
namely the resolution of singularities, in particular of the
quotient singularities. For this there is a well established set of
results that were all obtained by the mathematical community at the
beginning of 1990.s,  under the stimulus of string and supergravity
theory.
\par
First of all we fix sum vocabulary.
\begin{definizione}
The \textbf{canonical line bundle} $K_\mathbb{V}$ over a complex
algebraic variety $\mathbb{V}$ of complex dimension $n$ is the
bundle of  holomorphic $(n,0)$-forms $\Omega^{(n,0)}$ defined over
$\mathbb{V}$.
\end{definizione}
\begin{definizione}
An orbifold $\mathbb{V}/\Gamma$ of an algebraic variety modded by
the action of a finite group is named \textbf{Gorenstein} if the
isotropy subgroup $H_p\subset \Gamma$ of every point $p\in
\mathbb{V}$ has a trivial action on the canonical bundle
$K_\mathbb{V}$.
\end{definizione}
\begin{definizione}
A resolution of singularities $\pi \, : \,\,\mathbb{W}\to \,
\mathbb{X}\equiv \mathbb{V}/\Gamma$ is named \textbf{crepant}, if
$K_\mathbb{W}\, = \, \pi^\star K_\mathbb{X}$. In particular this
implies that the first Chern class of the resolved variety vanishes
($c_1\left( T\mathbb{W}\right)\, = \, 0$), if it vanishes for the
orbifold, namely if $c_1\left( T\mathbb{X}\right)\, = \, 0$.
\end{definizione}
In the case $\mathbb{V} \, = \, \mathbb{C}^n$, a resolution of
quotient singularity:
\begin{equation}
\pi \, : \,\,\mathbb{W}\to \, \mathbb{C}^n/\Gamma
\end{equation}
is crepant if the resolved variety $\mathbb{W}$ has vanishing first Chern class, namely if it is a Calabi-Yau
$q$--fold.
\par
The Gorenstein condition plus the request that there should be a
crepant resolution restricts the possible $\Gamma$.s to be subgroups
of $\mathrm{SL(n,\mathbb{C})}$.
\par
Concerning the crepant resolution of Gorenstein singularities $\mathbb{C}^n / \Gamma $, what was established
in the early 1990s is the following:
\begin{enumerate}
 \item For $n=2$ the classification of  Gorenstein singularities boils down to the classification of finite
 Kleinian subgroups $\Gamma \subset \mathrm{SU(2)}$. This latter is just the A-D-E classification and the
 crepant resolution of singularities is done in one stroke by the Kronheimer construction of
 ALE-manifolds\cite{kro1,kro2} via an HyperK\"ahler quotient of a flat HyperK\"ahler manifold
 $\mathbb{H}_\Gamma$, whose dimension and structure depends on the group $\Gamma$.
  \item For $n=3$ the classification of finite subgroups of $\mathrm{SL(3,\mathbb{C})}$ was performed at the
  very beginning of the XX century\cite{blicfeltus,millerus} and it is summarized in \cite{roanno}. As
  stressed by Markushevich in \cite{marcovaldo} in that list there are only two types of groups, either
  solvable groups or the simple group PSL(2,7) of order 168. For this reason the same Markushevich studied
  the resolution of the Gorenstein orbifold:
      \begin{equation}\label{miotesoro}
        \mathcal{O}_{168} \,  \equiv \, \frac{\mathbb{C}^3}{\mathrm{PSL(2,\mathbb{Z}_7)}}
      \end{equation}
      which corresponds to a unique truely new case. We are going to add several other physical motivations
      for the study of orbifolds with respect to
      \begin{equation}\label{L168}
        \mathrm{L_{168}} \, \equiv \, \mathrm{PSL(2,\mathbb{Z}_7)}
      \end{equation}
      or one of its maximal subgroups.  We postpone such discussion to later publications.
      Here we focus on the general form of $d=3$ Chern Simons gauge theories.
  \item For $n>3$ essentially nothing is known with the exception of those cases that can be reduced to
  singularities in $n=2,3$.
\end{enumerate}
\section{Rheonomic construction of matter coupled $\mathcal{N}=2$ gauge theories in $D=3$}
Following the results of \cite{Fabbri:1999ay} and \cite{Fabbri:1999hw} in this section we consider the
general form of a matter coupled Maxwell-Chern Simons gauge theory in $D=3$ space-time dimensions. In the
quoted references the K\"ahler manifold spanned by the Wess-Zumino multiplets was considered to be flat and
the action of the gauge group on the same was taken to be linear. In the present paper we need to be more
general. Hence the formulae of \cite{Fabbri:1999ay} and \cite{Fabbri:1999hw} here are geometrically rewritten
in terms of a generic K\"ahler metric, of Killing vectors and holomorphic moment maps. Supersymmetry is
$\mathcal{N}\, = \, 2$ in $D=3$ which amounts to the same as $\mathcal{N}=1$ in $D=4$. Just  as in
\cite{Fabbri:1999ay} and \cite{Fabbri:1999hw} we follow the off-shell approach with auxiliary fields which,
in the the final step of the construction,  can be eliminated through their own (algebraic) equation of
motion leading to the final form of the interactions among physical degrees of freedom.
\par
Furthermore, for all the reasons advocated in the introduction we are interested in the rheonomic
construction \footnote{For an overview of the rheonomic approach see the books \cite{castdauriafre} and
\cite{pietrograv}. In particular a relatively short modern presentation of the rheonomic principles is
presented in chapter 6 of the second volume of \cite{pietrograv}.} of the theory and in particular in the
explicit form of the rheonomic lagrangian which was indeed derived in \cite{Fabbri:1999ay} and
\cite{Fabbri:1999hw}. Here that result is generalized to arbitrary K\"ahler manifolds and to groups with an
arbitrary isometric holomorphic action.
\par
We start by fixing our conventions for the geometry of rigid
superspace.
\subsection{The supergeometry of $D=3$, $\mathcal{N}$=2 rigid superspace}
\label{supergeo}
$D=3,~\mathcal{N}$--extended superspace is viewed as the following
supercoset manifold:
\begin{equation}
{\cal M}^{\mathcal{N}}_3=\frac{\mathrm{ISO(1,2|\mathcal{N})}}
{\mathrm{SO(1,2)} }\, \equiv \, \frac{Z\left[
\mathrm{ISO(1,2|\mathcal{N})}\right ] }{\mathrm{SO(1,2)} \times
\mathbb{R}^{\mathcal{N}(\mathcal{N}-1)/2}}
\end{equation}
where $\mathrm{ISO(1,2\vert \mathcal{N})}$ is the
$\mathcal{N}$--extended Poincar\'e supergroup in three--dimensions.
Its superalgebra is the Inon\"u-Wigner contraction  of the
superalgebra $\mathrm{Osp(\mathcal{N}\vert 4)}$ spanned by the
generators $J_m$, $P_m$, $q^i$. The central extension $Z\left[
\mathrm{ISO(1,2|\mathcal{N})}\right ]$, which is not contained in
the contraction of  $\mathrm{Osp(\mathcal{N}\vert 4)}$, is obtained
by adjoining to $\mathrm{ISO(1,2|\mathcal{N})}$ the central charges
that generate the subalgebra
$\mathbb{R}^{\mathcal{N}(\mathcal{N}-1)/2}$. Specializing our
analysis to the case $\mathcal{N}\!\!=\!\!2$, we define the new
generators:
\begin{equation}
\left\{\begin{array}{ccl}
Q&=&\sqrt{2}q^-=(q^1-iq^2)\\
Q^c&=&\sqrt{2}iq^+=i(q^1+iq^2)\\
Z&=&Z^{12}
\end{array}\right.
\end{equation}
The left invariant one--form $\Omega$ on ${\cal M}^{\mathcal{N}}_3$
is the following object:
\begin{equation}
\Omega=e^mP_m-\ft{1}{2}\omega^{mn}J_{mn}+\overline{\psi^c}Q-\overline{\psi}Q^c
+{ \mathfrak{B}}Z\,.
\end{equation}
The superalgebra $\mathrm{ISO(1,2\vert N)}$ defines  all the structure constants apart from those relative to
the central charge that are trivially determined. Hence we can write:
\begin{eqnarray}
d\Omega-\Omega\wedge\Omega&=&\left(de^m-\omega^m_{\ n}\wedge e^n
+i\overline{\psi}\wedge\gamma^m\psi
+i\overline{\psi}^c\wedge\gamma^m\psi^c\right)P_m\nonumber\\
&&-\ft{1}{2}\left(d\omega^{mn}-\omega^m_{\ p}\wedge\omega^{pn}\right)J_{mn}\nonumber\\
&&+\left(d\overline{\psi}^c
+\ft{1}{2}\omega^{mn}\wedge\overline{\psi}^c\gamma_{mn}\right)Q\nonumber\\
&&+\left(d\overline{\psi}
-\ft{1}{2}\omega^{mn}\wedge\overline{\psi}\gamma_{mn}\right)Q^c\nonumber\\
&&+\left(d{ \mathfrak{B}}+i\overline{\psi}^c\wedge\psi^c
-i\overline{\psi}\wedge\psi\right)Z
\end{eqnarray}
Imposing the Maurer-Cartan equation $d\Omega-\Omega\wedge\Omega=0$
is equivalent to imposing flatness in superspace, i.e. global
supersymmetry. So we have
\begin{eqnarray}
0 &=& T^m \, \equiv \, \mathcal{D}e^m \, + \, {\rm i}\,
\left(\bar{\psi}_c \, \wedge \, \gamma^m \, \psi_c \, +
\, \bar{\psi} \, \wedge \, \gamma^m \, \psi\right ) \nonumber\\
0 &=& R^{mn}\, \equiv \,d\omega^{mn} \, - \, \omega^{mp} \, \wedge \, \omega^{pn} \nonumber \\
0 &=& \rho \, \equiv \, d\psi \, + \, \frac{1}{2} \, \omega^{mn} \, \wedge \, \gamma_{mn} \, \psi \nonumber\\
0 &=& \rho_c \, \equiv \, d\psi_c \, - \, \frac{1}{2} \, \omega^{mn} \, \wedge \, \gamma_{mn} \, \psi_c \nonumber\\
0  &=& RZ   \, \equiv \, d\mathfrak{B} \, + \, {\rm i}\,
\left(\bar{\psi}_c \, \wedge \, \psi_c \, - \, \bar{\psi} \, \wedge
\, \psi\right ) \label{struceque}
\end{eqnarray}
The above equations are nothing else but  the statement that the
curvatures of the $\mathcal{N}=2$ super-Poincar\'e group are zero.
The simplest solution for the supervielbein and for the connection
satisfying the above structural equations of rigid superspace is the
following one:
\begin{eqnarray}
  e^m &=& dx^m\, - \,  {\rm i}\, \left(\bar{\theta}_c \, \gamma^m \, d\theta_c \, + \, \bar{\theta} \,  \gamma^m
  \, d\theta\right ) \quad \nonumber\\
  \omega^{mn} &=& 0\nonumber\\
  \psi_c &=& d\theta_c \nonumber\\
  \psi &=& d\theta \nonumber\\
  \mathfrak{B} &=& - \, {\rm i}\, \left(\bar{\theta}_c \, d\theta_c \, - \, \bar{\theta} \,   d\theta\right )
  \label{superspazio}
\end{eqnarray}
where $x^m$ are the standard coordinates of flat Minkowsky space and $\theta$ are the anticommuting
grassmanian supercoordinates. They form a $D=3$ Dirac spinor corresponding to four independent components.
Our conventions for the $D=3$ Dirac matrices are the following ones:
\begin{equation}\label{convenzie}
    \begin{array}{ccccccc}
       \gamma^0 & = & - \,{\rm i} \, \sigma^2 & ; & \eta_{mn} & =  & \mbox{diag} \,\left(-,+,+\right) \\
       \gamma^1 & = & \sigma^3 & ; & C\, \gamma^m \, C^{-1} & = & - \, \left(\gamma^m\right)^T\\
       \gamma^2 & = & \sigma^1 & ; & \gamma^{mn} & = & \frac{1}{2} \, \left[\gamma^m \, , \, \gamma^n\right]
     \end{array}
\end{equation}
The superderivatives
\begin{equation}
\left\{\begin{array}{ccl}
D_m&=&\partial_m\\
D&=&\frac{\partial}{\partial\overline{\theta}}-i\gamma^m\theta\partial_m\\
D^c&=&\frac{\partial}{\partial\overline{\theta}^c}-i\gamma^m\theta^c\partial_m\\
\end{array}\right.,
\end{equation}
are the vector fields dual to the above one--forms.
\par
Let us observe that  by $\psi_c$ we denote the conjugate of the
spinor $\psi$ according to the following definition in terms of the
charge conjugation matrix:
\begin{equation}\label{coniugatus}
    \psi_c \, \equiv \, C \, \bar{\psi}^T \quad ; \quad \bar{\psi} \, = \, \psi^\dagger \, \gamma^0 \quad ;
    \quad C \, = \, {\rm i} \, \sigma^2
\end{equation}
\paragraph{Relevant Fierz Identities.} Furthermore in the further development of the
theory some Fierz identities are particularly useful and relevant:
\begin{eqnarray}
  \overline{\psi} \wedge \gamma_m \,\psi  &=& \overline{\psi}_c \wedge \gamma_m \,\psi_c \label{fierz1}\\
  \overline{\psi} \wedge \psi  &=& - \, \overline{\psi}_c \wedge \psi_c \\
  \overline{\psi} \wedge \psi_c &=& 0
\end{eqnarray}
%
%
\subsection{The ingredients} As stated in the
introduction we are interested in the general form of an
$\mathcal{N}=2,~d=3$ super Yang Mills theory coupled to $n$ chiral
multiplets arranged into a generic representation $\cal R$ of the
gauge group $\cal G$.
\par
In $\mathcal{N}=2,~d=3$ supersymmetric theories, two formulations
are allowed: the on--shell and the off--shell one. In the on--shell
formulation which contains only the physical fields, the
supersymmetry transformations rules close the supersymmetry algebra
only upon use of  the field equations. On the other hand the
off--shell formulation contains further auxiliary, non dynamical
fields that make it possible for the supersymmetry transformations
rules to close the supersymmetry algebra identically. By solving the
field equations of the auxiliary fields these latter can be
eliminated   and the on--shell formulation is thus retrieved. We
adopt the off--shell formulation.
%
%
\subsubsection{The gauge multiplet}
The three--dimensional $\mathcal{N}=2$ vector multiplet contains the
following Lie-algebra valued fields:
\begin{eqnarray}\label{vectorm}
\mbox{vect. mult.} & = &\left\{
\underbrace{\mathcal{A}^\Lambda}_{\mbox{gauge one-form}} \, , \,
    \underbrace{\lambda^\Lambda \, , \, \lambda^\Lambda_c}_{\mbox{gauginos}} \, ,
    \, \underbrace{M^\Lambda}_{\mbox{phys. scalar}} \, \underbrace{D^\Lambda}_{\mbox{aux. scalar}} \right\}
\end{eqnarray}
where ${\cal A}={\cal A}^\Lambda t_\Lambda$ is the real gauge
connection one--form, $\lambda^\Lambda$ and $\lambda^\Lambda_c$ are
two complex Dirac spinors (the \emph{gauginos}),  $M^\Lambda$ and
$D^\Lambda$ are real scalars, the latter being the auxiliary field.
The capital Greek indices $\Lambda,\Sigma,\dots$ span the adjoint
representation of the gauge group $\mathcal{G}$.
\par
The field strength $2$-form is defined below:
\begin{eqnarray}\label{defF}
\mathfrak{F}&\equiv & d{\cal A}+{\cal A}\wedge{\cal A}\,=
\,\mathfrak{F}^\Lambda \, t_\Lambda\nonumber\\
\mathfrak{F}^\Lambda & = & \mathrm{d}\mathcal{A}^\Lambda \,
     + \, f^{\Lambda}_{\phantom{\Lambda}\Delta\Sigma} \,
    \mathcal{A}^\Delta \, \wedge \, \mathcal{A}^\Sigma
\end{eqnarray}
The covariant derivative on any other  other field $X$ of the gauge
multiplet is defined below:
\begin{equation}\label{defnablagauge}
\nabla X=dX+\left[{\cal A},X\right]\,.
\end{equation}
From (\ref{defF}) and (\ref{defnablagauge}) we obtain the
Bianchi identity:
\begin{equation}
\nabla^2 X=\left[\mathfrak{F},X\right]\,.
\end{equation}
The vector multiplets contains $4_B\oplus 4_F$ bosonic and fermionic
off--shell degrees of freedom for each generator of the gauge group.
\par
The off-shell  rheonomic parametrization of the vector multiplet
\emph{curvatures}, consistent with the Bianchi identities is given
below:
\begin{eqnarray}
  \mathfrak{F}^\Lambda \, &=& F^\Lambda_{mn} \, e^M \, \wedge \, e^n \, - \, {\rm i} \bar{\psi}_c \,
  \gamma_m \, \lambda^\Lambda \, \wedge \, e^m \, - \, {\rm i} \bar{\psi} \, \gamma_m \, \lambda^\Lambda_c \,
  \wedge \, e^m \nonumber\\
  &&  - \, {\rm i} M^\Lambda \left( \overline{\psi}\wedge\psi - \overline{\psi}_c\wedge\psi_c\right)\nonumber\\
  \nabla \lambda^\Lambda \, \equiv \,   \mathrm{d}\lambda^\Lambda \, + \, \left[ \mathcal{A} \, ,
  \, \lambda\right]^\Lambda &=&
  \nabla_m \lambda^\Lambda \, e^m \, + \, \nabla_mM^\Lambda \, \gamma^m \, \psi_c \, - \, F^\Lambda_{mn} \,
  \gamma^{mn} \, \psi_c \, + \, {\rm i} D^\Lambda \, \psi_c \nonumber\\
\nabla \lambda^\Lambda_c \, \equiv \,   \mathrm{d}\lambda^\Lambda_c
\, + \, \left[ \mathcal{A} \, , \, \lambda_c\right]^\Lambda &=&
  \nabla_m \lambda^\Lambda_c \, e^m \, - \, \nabla_mM^\Lambda \, \gamma^m \, \psi \, - \, F^\Lambda_{mn} \,
  \gamma^{mn} \, \psi \, - \, {\rm i} D^\Lambda \, \psi \nonumber\\
\nabla M^\Lambda\, \equiv \,   \mathrm{d}M^\Lambda \, + \, \left[
\mathcal{A} \, , \, M\right]^\Lambda &=&
  \nabla_m M^\Lambda_c \, e^m \, + \, {\rm i} \,  \bar{\psi} \, \lambda^\Lambda_c \,  - \, {\rm i} \bar{\psi}_c \,
  \lambda^\Lambda  \nonumber\\
  \nabla D^\Lambda\, \equiv \,   \mathrm{d}D^\Lambda \, + \, \left[ \mathcal{A} \, , \, D\right]^\Lambda &=&
  \nabla_m D^\Lambda_c \, e^m \, + \,   \bar{\psi} \, \gamma^m \, \nabla_m \,\lambda^\Lambda_c \,  - \,  \bar{\psi}_c \,
  \gamma^m \, \nabla_m \, \lambda^\Lambda \nonumber\\
   &&\, - \, {\rm i} \, \bar{\psi }\,\ \left[\lambda_c \, , \, M \right]^\Lambda \, - \, {\rm i} \, \bar{\psi }_c\,\
   \left[\lambda \, , \, M \right]^\Lambda \label{rheovector}
\end{eqnarray}
and we also have:
\begin{equation}
\left\{\begin{array}{ccl}
\nabla F_{mn}&=&V^p\nabla_pF_{mn}+i\overline{\psi}^c\gamma_{[m}\nabla_{n]}\l
+i\overline{\psi}\gamma_{[m}\nabla_{n]}\l^c\\
\nabla\nabla_mM&=&V^n\nabla_n\nabla_mM+i\overline{\psi}\nabla_m\l^c
-i\overline{\psi}^c\nabla_m\l+\overline{\psi}^c\gamma_m\left[\l,M\right]
+\overline{\psi}\gamma_m\left[\l^c,M\right]\\
\nabla\nabla_m\l&=&V^n\nabla_n\nabla_m\l+\nabla_m\nabla_nM\gamma^n\psi^c
-\nabla_mF_{np}\gamma^{np}\psi^c\\
&&+i\nabla_mP\psi^c+\overline{\psi}\gamma_m\left[\l^c,\l\right]\\
\nabla_{[p}F_{mn]}&=&0\\
\nabla_{[m}\nabla_{n]}M&=&i\left[F_{mn},M\right]\\
\nabla_{[m}\nabla_{n]}\l&=&i\left[F_{mn},\l\right]
\end{array}\right.
\end{equation}
\subsubsection{The chiral Wess Zumino multiplets}
The Wess Zumino multiplets have the same structure in $D=3$ as they
have in $D=4$.
\begin{equation}\label{vecmult}
    \mbox{WZ. mult.} \, = \, \left\{ \underbrace{z^i}_{\mbox{complex scalars}} \, , \,
    \underbrace{\chi^i \, , \, \chi^i_c}_{\mbox{chiralinos}} \, , \, \underbrace{\mathcal{H}^i}_{\mbox{complex aux fields}}
    \, \right\}
\end{equation}
The complex scalar fields $z^i$ parameterize a K\"ahler manifold
$\mathcal{M}_K$ whose geometry is determined by a K\"ahler potential
$\mathcal{K}\left(z,{\bar z}\right)$ yielding as usual the metric:
\begin{equation}\label{Kalmetra}
    g_{ij^\star} \, = \, \partial_i \,\partial_{j^\star} \, \mathcal{K}
\end{equation}
The continuous isometries (if any) of this metric are generated by
holomorphic Killing vectors $k_\Lambda^i(z)$ according to:
\begin{equation}\label{isometrala}
    z^i \, \mapsto \, z^i \, + \, \epsilon^\Lambda \, k_\Lambda^i(z)
\end{equation}
and the vector multiplets can be used to gauge  such symmetries and
make them local. One sets:
\begin{eqnarray}
  \nabla z^i  &\equiv& \mathrm{d}z^i  \, + \, \mathcal{A}^\Lambda \, k_\Lambda^i(z) \nonumber\\
  \nabla \chi^i  &\equiv& \mathrm{d}\chi^i  \, + \, \hat{\Gamma}^i_{\phantom{j}j} \, \chi^j \quad\quad\quad ;
  \quad \hat{\Gamma}^i_{\phantom{j}j} \,\,\, = \, {\Gamma}^i_{\phantom{j}j} \, + \, \mathcal{A}^\Lambda \,
  \partial_j k_\Lambda^i \nonumber \\
  \nabla \chi^{j^\star}  &\equiv& \mathrm{d}\chi^{j^\star}  \, + \, \hat{\Gamma}^{j^\star}_{\phantom{j^\star}k^\star}
  \, \chi^{k^\star} \quad ; \quad \hat{\Gamma}^{j^\star}_{\phantom{j^\star}k^\star}={\Gamma}^{j^\star}_{\phantom{j^\star}k^\star}\,
  +\mathcal{A}^\Lambda \, \partial_j k_\Lambda^i\nonumber\\
  \nabla \mathcal{H}^i  &\equiv& \mathrm{d}\mathcal{H}^i   \,
   + \, \hat{\Gamma}^i_{\phantom{j}j} \, \mathcal{H}^j \label{WZcovder}
\end{eqnarray}
If one compares the above equations with the similar ones that appear in the coupling of chiral Wess Zumino
multiplets to $\mathcal{N}=1$, $D=4$ supergravity (see for instance \cite{primosashapietro,piesashatwo}), one
should notice the absence of the K\"ahler connection $\mathcal{Q}$ in the covariant derivative of the
chiralinos. This is the main structural difference between the case of local and rigid supersymmetry. In the
second case, which is that of interest to us in the present paper, there is no $\mathrm{U(1)}$-bundle over
the K\"ahler manifold which is not requested to be Hodge-K\"ahler, but simply K\"ahler. Correspondingly the
fermions do not transform as sections of the Hodge bundle and there is no Hodge-K\"ahler connection in their
covariant derivative. We have only the gauged version $\hat \Gamma$ of the Levi-Civita holomorphic connection
since the chiralinos transform as sections of the tangent bundle $T\mathcal{M}_K$.
\par
The off--shell rheonomic parameterizations of the chiral multiplet
fields are the following ones:
\begin{eqnarray}
\nabla z^i \, &=&
  \nabla_m z^i \, e^m \, + \, 2\, \bar{\psi}_c \,\chi^i \nonumber\\
\nabla \bar{z}^{i ^\star}\, &=&
  \nabla_m \bar{z}^{i ^\star}\, e^m \, + \, 2\, \bar{\psi} \,\chi^{i^\star} _c\nonumber\\
\nabla \chi^i &=&
\nabla_m \chi^i\, e^m \, - \, {\rm i} \,  \nabla_m  \, z^i \,
\gamma^m \, \psi_c  \, + \, \mathcal{H}^i \,\psi \, - \, {\rm i} \,
M^\Lambda \, k^i_\Lambda \, \psi_c \nonumber\\
\nabla \chi^{j^\star} &=& \nabla_m \chi^{j^\star}\, e^m \, + \, {\rm
i} \,  \nabla_m  \, \bar{z}^{j^\star} \, \gamma^m \, \psi  \, - \,
\bar{\mathcal{H}}^{j^\star} \,\psi_c \, + \, {\rm i} \,M^\Lambda \,
k^i_\Lambda \, \psi \label{rheovectorChi}
\end{eqnarray}
and from them we read off the supersymmetry transformation rules.
\par
Additional essential items in the construction of the theory are the
moment maps defined by the following equation:
\begin{equation}\label{momentimappi}
k^i_\Lambda \, = \, {\rm i} \,g^{ij^\star} \,\partial_{j^\star}
\mathcal{P}_\Lambda \quad ; \quad k^i_\Lambda \, = \, - \, {\rm i}
\,g^{ij^\star} \,\partial_{j^\star} \mathcal{P}_\Lambda
\end{equation}
\subsection{The rheonomic lagrangian}
Using the rules of the rheonomic approach, the rheonomic lagrangian of a general matter coupled gauge theory
in $D=3$ was determined in \cite{Fabbri:1999ay} for the case where the K\"ahler manifold $\mathcal{M}_K$ is
flat and the action of the gauge group on the chiral multiplets is linear. The transcription of that result
to the general case of an arbitrary $\mathcal{M}_K$ is rather straightforward and actually gives rise to more
compact and more elegant formulae. We report the result of \cite{Fabbri:1999ay} in its generalized form.
\par
The rheonomic Lagrangian can be organized in the following way:
\begin{eqnarray}\label{rheolagga}
    \mathcal{L}_{rheo}^{\mathcal{N}=2}& = &{\cal L}^{gauge}_{rheo}\,+ \,{\cal
    L}^{chiral}_{rheo}\nonumber\\
{\cal L}^{gauge}_{rheo}&=&{\cal L}_{rheo}^{Maxwell} +{\cal
L}_{rheo}^{Chern-Simons} +{\cal L}_{rheo}^{Fayet-Iliopoulos}\nonumber\\
{\cal L}^{chiral}_{rheo}&=&{\cal L}_{rheo}^{Kahler}+{\cal
L}_{rheo}^{superpotential}
\end{eqnarray}
Next we display the various addends mentioned in
eq.(\ref{rheolagga}). We begin with the kinetic part of the gauge
lagrangian, named after Maxwell.
\begin{eqnarray}\label{MM1}
{\cal L}_{rheo}^{Maxwell}&=& \mathbf{e} \,
\mathrm{Tr}\left\{-F^{mn}\left[\mathfrak{F}
+i\overline{\psi}^c\gamma_m\l\wedge e^m+i\overline{\psi}\gamma_m\l^c
\wedge e^m
-2iM\overline{\psi}\wedge\psi\right]\wedge e^p\epsilon_{mnp}\right.\nonumber\\
&&\left.+\ft{1}{6}F_{qr}F^{qr}e^m\wedge e^n \wedge e^p\epsilon_{mnp}
-\ft{1}{4}i\epsilon_{mnp}\left[\nabla\overline{\l}\gamma^m\l
+\nabla\overline{\l}^c\gamma^m\l^c\right]\wedge
e^n \wedge e^p\right.\nonumber\\
&&\left.\ft{1}{2}\epsilon_{mnp}\Phi^m\left[\nabla
M-i\overline{\psi}\l^c +i\overline{\psi}^c\l\right]\wedge e^n \wedge
e^p
-\ft{1}{12}\Phi^d\Phi_d \, \epsilon_{mnp}e^m\wedge e^n \wedge e^p\right.\nonumber\\
&&\left. +\nabla M \wedge \overline{\psi}^c\gamma_c\l \wedge e^p
-\nabla M \wedge \overline{\psi}\gamma_p\l^c \wedge e^p\right.\nonumber\\
&&\left. +\mathfrak{F}\wedge \overline{\psi}^c\l+\mathfrak{F}\wedge
\overline{\psi}\l^c
+\ft{1}{2}i\overline{\l}^c\l\overline{\psi}^c\wedge \gamma_m\psi
\wedge  e^m
+\ft{1}{2}i\overline{\l}\l^c\overline{\psi}\wedge \gamma_m\psi^c \wedge e^m\right.\nonumber\\
&&\left.+\ft{1}{12}{ D}^2 \, e^m \wedge e^n \wedge e^p\epsilon_{mnp}
-2i(\overline{\psi}\wedge \psi)\, M\, \wedge
\left[\overline{\psi}^c\l
+\overline{\psi}\l^c\right]\right.\nonumber\\
&&\left.-\ft{1}{6}M\left[\overline\l,\l\right]e^m \wedge e^n\wedge
e^p \, \e_{mnp}\right\}
\end{eqnarray}
The Chern-Simons part of the vector multiplet rheonomic lagrangian
is instead the following one:
\begin{eqnarray}\label{MM2}
{\cal L}_{rheo}^{Chern-Simons}&=&\alpha \, \mathrm{Tr}\left\{
-\left(\mathcal{A}\wedge \mathfrak{F}\,+\,\ft 23 \,\mathcal{A}\wedge
\mathcal{A}\wedge \mathcal{A}\right)
-\ft{1}{3}M\, D \, e^m \wedge e^n \wedge  e^p\, \epsilon_{mnp}\right.\nonumber\\
&+&\ft{1}{3}\left(\overline{\lambda}\lambda +
\overline{\lambda}_c\lambda_c\right) \,e^m \wedge e^n \wedge e^p \,
\epsilon_{mnp} +M\left[\overline{\psi}^c\gamma^m\lambda \,
-\overline{\psi}\gamma^m\lambda^c\right]\wedge e^n \wedge e^p \epsilon_{mnp}\nonumber\\
&&\left.-2iM^2\overline{\psi}\wedge\gamma_m\psi \wedge e^m\right\}
\end{eqnarray}
while the Fayet-Iliopoulos addend has the following appearance
\begin{eqnarray}\label{MM3}
{\cal L}_{rheo}^{Fayet-Iliopoulos}&=&\zeta^I \,\mathrm{Tr}\left\{
\mathfrak{C}_I\left[ -\ft{1}{6}D \, e^m \wedge e^n \wedge e^p \,
\e_{mnp} +\ft{1}{2}\left(\overline{\psi}^c\gamma^m\l
-\overline{\psi}\gamma^m\l^c\right)\wedge e^n \wedge e^p \e_{mnp}
\right.\right.\nonumber\\
&&\left.\left.-2iM\, \overline{\psi}\wedge\gamma_m\psi \wedge e^m
-2i{\cal A}\wedge \overline{\psi}\wedge \psi\right]\right\}
\end{eqnarray}
where $\mathfrak{C}_I$ denotes a basis of generators of the center
$Z\left[\mathbb{G}\right]$ of the gauge Lie algebra $\mathbb{G}$.
\par
The two sectors of the chiral rheonomic lagrangian are displayed
next.
\begin{eqnarray}\label{MM4}
{\cal L}_{rheo}^{Kahler}&=&\ft{1}{2}\,\overline{\Pi}^{m\,i^*}
\eta_{i^*j}\left[\nabla z^j
-2\overline{\psi}^c\chi^j\right]\wedge e^n \wedge e^p \, \e_{mnp}\nonumber\\
&+&\ft{1}{2}\, \Pi^{m\,i}\eta_{ij^*}\left[\nabla \overline z^{j^*}
-2\overline{\chi}\psi^{c\,j^*}\right] \wedge e^n \wedge e^p \, \e_{mnp}\nonumber\\
&-&\ft{1}{6}\, \eta_{ij^*}\Pi_q^{\,i}
\overline{\Pi}^{q\,j^*}e^m \wedge e^n \wedge e^p \, \e_{mnp}\nonumber\\
&+&\ft{1}{2}i
\eta_{ij^*}\left[\overline{\chi}^{j^*}\gamma^m\nabla\chi^i
+\overline{\chi}^{c\,i}\gamma^m\nabla\chi^{c\,j^*}
\right]\wedge e^n \wedge e^p \, \e_{mnp}\nonumber\\
&+&2i\eta_{ij^*}\left[\nabla z^i \wedge
\overline{\psi}\gamma_m\chi^{c\,j^*} -\nabla \overline z^{j^*}
\wedge \overline{\chi}^{c\,i}\gamma_m\psi
\right] \wedge e^m\nonumber\\
&-&2i\eta_{ij^*}\left(\overline{\chi}^{j^*}\gamma_m\chi^i\right)
\left(\overline{\psi}^c \wedge \psi^c\right)\wedge e^m
-2i\eta_{ij^*}\left(\overline{\chi}^{j^*}\chi^i\right)
\left(\overline{\psi}^c \wedge \gamma_m\psi^c\right) \wedge e^m\nonumber\\
&+&\ft{1}{6}\eta_{ij^*}\mathcal{H}^i\overline{\mathcal{H}}^{j^*}\,
e^m \wedge e^n \wedge e^p \e_{mnp} +\left(\overline{\psi}\wedge
\psi\right) \wedge \eta_{ij^*}\left[\overline z^{j^*}\nabla z^i
-z^i\nabla \overline z^{j^*}\right]\nonumber\\
&+&i \, g_{ij^\star}\, M^\Lambda \, k_\Lambda^i
\overline{\chi}^{j^*}\gamma^m\psi^c \wedge
e^n \wedge e^p \e_{mnp} \nonumber\\
&+&i \, g_{i^\star j}\, M^\Lambda \, k_\Lambda^{i^\star}
\overline{\chi}^{j}\gamma^m\psi \wedge
e^n \wedge e^p \e_{mnp}\nonumber\\
&+& \left(- \, \frac{1}{3} \, M^\Lambda \,\left(
\partial_i k^j_\Lambda \, g_{j\ell^\star} \, \bar{\chi}^{\ell^\star}
\,\chi^i \,+ \,
\partial_{i^\star} k^{j^\star}_\Lambda \, g_{j\ell^\star}
\bar{\chi}^\ell_c \, \chi^{i^\star}_c \right)\right. \nonumber\\
&&+\left.\, {\rm i}\, \frac{1}{3} \, \left( \bar{\chi}^{j^\star}_c
\, \lambda^\Lambda \, k^i_\Lambda \, - \, \bar{\chi}^{i}_c \,
\lambda^\Lambda \, k^{j^\star}_\Lambda\right) \, g_{ij^\star}
\,\right.\nonumber\\
&&\left. + \ft 16 D^\Lambda \, \mathcal{P}_\Lambda(z,\bar{z}) \, -
\, \frac{1}{6} M^\Lambda \, M^\Sigma \, k^i_\Lambda \,
k_\Sigma^{j^\star} \, g_{ij^\star} \right)\,
e^m \wedge e^n \wedge e^p \, \e_{mnp}\nonumber\\
&-& \,\ft{1}{2}\, \mathcal{P}_\Lambda(z,\bar{z}) \left(
\,\overline{\psi}_c\gamma^m\l^\Lambda -
\overline{\psi}\gamma^m\l^\Lambda_c \right) \, \wedge e^n\wedge e^p \, \e_{mnp}\nonumber\\
&+&2iM^\Lambda \mathcal{P}_\Lambda(z,\bar{z})\,
\overline{\psi}\wedge \gamma_m\psi \wedge e^m
\end{eqnarray}
\begin{eqnarray}\label{MM5}
{\cal L}_{rheo}^{superpotential}&=& - 2i\left[W(z)+\overline W(\overline z)\right]
\overline{\psi}\wedge \gamma_m\psi_c \wedge e^m \nonumber\\
&-&i\left[\partial_{j^*}\overline W(\overline z) \, \overline{\chi}^{j^*}\gamma^m \psi+\partial_jW
({z})\overline{\chi}^{j}_c\gamma^m
\psi^c \right]\wedge e^n\wedge e^p \,\e_{mnp}\nonumber\\
&+&\ft{1}{6}\left[\partial_i\partial_jW(z)\overline{\chi}^{i}_c\chi^j +\partial_{i^*}\partial_{j^*}\overline
W(\overline z)
\overline{\chi}^{i^*}\chi^{j^*}_c\right]\, e^m \wedge e^n \wedge e^p \, \e_{mnp}\nonumber\\
&-&\ft{1}{6}\left[\mathcal{H}^i\partial_iW(z)+{ \mathcal{H}}^{j^*}\partial_{j^*} \overline W(\overline
z)\right]e^m\wedge e^n \wedge e^p \, \e_{mnp}
\end{eqnarray}
%
%
\section{The space--time Lagrangian of the Maxwell-Chern-Simons theory and some of its applications}
In the rheonomic approach (\cite{castdauriafre}), the total
three--dimensional $\mathcal{N}\!\!=\!\!2$ rheonomic lagrangian:
\begin{equation}
{\cal L}^{\mathcal{N}=2}_{rheo}={\cal L}^{gauge}_{rheo}+{\cal
L}^{chiral}_{rheo}
\end{equation}
is a closed  three--form defined in superspace $d{\cal
L}^{N=2}_{rheo}=0$.


The space--time lagrangian,
\begin{equation}\label{N=2stLag}
{\cal L}^{\mathcal{N}=2}_{st}={\cal L}_{kinetic} +{\cal
L}_{2fermi}+{\cal L}_{potential}
\end{equation}
The first addend is the kinetic part admitting the following
explicit form:
\begin{eqnarray}
  \mathcal{L}_{kin}&=& - \, \alpha \, \mbox{Tr} \,\left(\mathfrak{F} \, \wedge \, \mathcal{A} \, + \, \frac{2}{3}\, \mathcal{A}\wedge \, \mathcal{A}\wedge \, \mathcal{A}\right)\, - \, \mathbf{e} \, \mbox{Tr} \left(F^{pq} \,\mathfrak{F}\right) \, \wedge \, e^r \, \epsilon_{pqr}\nonumber\\
  \null&&+\,\left( \frac{1}{2} \, g_{ij^\star} \, \left(\Pi^{m\mid i} \, \nabla \bar{z}^{j^\star} \, + \,  \bar{\Pi}^{m\mid j^\star} \, \nabla z^{i}\right) \, + \, \mathbf{e}\, \mbox{Tr} \left(\Phi^m \, \mathrm{d}M\right)\right)\, \wedge \, e^n \, \wedge \,  e^p \, \epsilon_{mnp}\nonumber\\
  \null && +\left(\mathbf{e}\,\frac{1}{6}F_{qr} \, F^{qr} \, - \, \frac{1}{6} \, \, g_{ij^\star} \, \Pi^{m\mid i} \, \bar{\Pi}^{m\mid j^\star} \, - \, \mathbf{e}\,\frac{1}{12} \, \Phi^p\, \Phi_p \right) e^m \,  \wedge \, e^n \, \wedge \,  e^p \, \epsilon_{mnp}\nonumber\\
 &&\left(\, \mathbf{e}\, \mbox{Tr}\left(\nabla \bar{\lambda}\, \gamma^m \lambda \, + \, \nabla \bar{\lambda}_c\, \gamma^m \lambda_c\right)  + {\rm i}\frac{1}{2} \, g_{ij^\star} \, \left(\bar{\chi}^{j^\star} \, \gamma^m \,\nabla \chi{i} \, + \,  \bar{\chi}^{i}_c \, \, \gamma^m \,
  \nabla \chi{i^\star}_c\right)\right) \, \wedge \, e^n \, \wedge \,  e^p \, \epsilon_{mnp} \nonumber\\
\label{kinosi}
\end{eqnarray}
In the above equation $\Pi^{m\mid i} $ and $\bar{\Pi}^{m\mid
j^\star}$ are zero forms with both a space-time vector index $m$ and
a K\"ahler manifold vector index $i$ (or $j^\star$). This denotes
that these $0$-forms transform both as sections of the tangent
bundle to space-time and as sections of the tangent bundle to the
K\"ahler manifold $T\mathcal{M}_K$. From variation in $\delta
\,\Pi^{m\mid i}$ and $\delta \,{\bar\Pi}^{m\mid i}$ we obtain:
\begin{equation}\label{piquipila}
    \Pi^{m\mid i} \, = \, \nabla_m z^i \quad ; \quad \bar{\Pi}^{m\mid j^\star } \, = \, \nabla_m \bar{z}^{j^\star}
\end{equation}
Similarly for the $0$-forms $\Phi^m_\Lambda$ which have an index in
the adjoint representation of the gauge group $\Lambda$ and a
space-time vector index  $m$. Their equation of motion is algebraic
and identifies them with the space-time derivatives of the
corresponding scalar fields $M^\Lambda$ belonging to the vector
multiplets:
 \begin{equation}\label{Fione}
    \Phi_m^\Lambda \, = \, \nabla_m M^\Lambda
 \end{equation}
Finally the $0$-form $F_{pq}$ with two antisymmetric space-time
indices has also an algebraic field equation and gets identified
with the space-time components of the Yang-Mills curvature:
\begin{equation}\label{forticampo}
    \mathfrak{F} \, = \, F_{pq} \, e^p \, \wedge \, e^q
 \end{equation}
\par
The parameter $\mathbf{e}\, = \, \frac{1}{g}$  where $g$ is the Yang-Mills gauge coupling constant, sits in
front of all the terms that compose the separately supersymmetric kinetic terms of the vector multiplets. The
parameter $\alpha$, instead, sits in front of all the terms that provide the supersymmetrization of the
Chern-Simons term. We find these parameters also in the other sectors of the lagrangian. Setting
$\mathbf{e}\, = \, 0$ we obtain a purely Chern Simons gauge theory, while setting $\alpha \, = \, 0$ we
suppress the Chern Simons term and all of its supersymmetric partners. In its most general form the
lagrangian contains three invariants: the kinetic term associated with $\mathbf{e}$, the Chern-Simons
associated with $\alpha$ and the Fayet Iliopoulos term which occurs only in the potential sector.
\par
The next part of the lagrangian is the $2$-fermi part that contains
the scalar--field--dependent mass--terms of the fermions:
\begin{eqnarray}
 \mathcal{L}_{2fermi}&=& \left( \, - \, \frac{1}{3} \, M^\Lambda \,
 \left( \partial_i k^j_\Lambda \, g_{j\ell^\star} \, \bar{\chi}^{\ell^\star}
\,\chi^i \,+ \,  \partial_{i^\star} k^{j^\star}_\Lambda \,
g_{j\ell^\star} \bar{\chi}^\ell_c \, \chi^{i^\star}_c \right) \,
 + \, \frac{\alpha}{3} \, \left( \bar{\lambda}^\Lambda \, \lambda^\Sigma \,
 + \, \bar{\lambda}^\Lambda_c \, \lambda^\Sigma_c\right) \, \mathbf{\kappa}_{\Lambda\Sigma}\right.\nonumber\\
&&\left. \, +\, {\rm i}\, \frac{1}{3} \, \left(
\bar{\chi}^{j^\star}_c \, \lambda^\Lambda \, k^i_\Lambda \, - \,
\bar{\chi}^{i}_c \, \lambda^\Lambda \, k^{j^\star}_\Lambda\right) \,
g_{ij^\star} \, - \, \mathbf{e} \, \frac{1}{6} \,\mbox{Tr} \left(M
\, \left(\left[\lambda\, , \, \lambda\right]
\, + \, \left[\lambda_c\, , \, \lambda_c\right] \right)\right)\right.\nonumber\\
  &&\left. + \, \frac{1}{6} \, \left( \partial_i\partial_j W\, \bar{\chi}^i_c \, \chi^j \,
  + \, \partial_{i^\star}\partial_{j^\star} {\overline{W}}\,
  \bar{\chi}^{i^\star} \, \chi^{j^\star}_c\right) \right) \, \epsilon_{mnp} \, e^m \, \wedge \, e^n \, \wedge \, e^p \label{duefermi}
\end{eqnarray}
In the above equation  $\mathbf{\kappa}_{\Lambda\Sigma}$ denotes the
Killing metric on the Lie algebra of the gauge group or any other
invariant metric if the Lie algebra is not semisimple. Furthermore
the holomorphic function $W(z)$ is the so called
\textit{superpotential} which, together with the K\"ahler metric,
determines all self-interactions of the chiral multiplets. Finally
both in equation (\ref{kinosi}) and the (\ref{duefermi}) the terms
with the parameter $\alpha$ in front are part of the Chern-Simons
term supersymmetrization. Together with the residual $\alpha$-terms
that we have in the next part of the Lagrangian, namely in the
potential part, those above constitute a separate supersymmetry
invariant. We can switch on and off the $\alpha$-terms preserving
off-shell supersymmetry of the Lagrangian.
\par
The next and last part of the space-time lagrangian is the potential
part. It has the following form:
\begin{eqnarray}
  \mathcal{L}_{pot}&=&-V\left(M,P,\mathcal{H},z,\bar{z}\right)  \, \epsilon_{mnp} \, e^m \, \wedge \, e^n \, \wedge \, e^p \nonumber\\
  V\left(M,P,\mathcal{H},z,\bar{z}\right)&=& \left(\frac{\alpha}{3} \,  M^\Lambda \,\mathbf{\kappa}_{\Lambda\Sigma}\,  - \, \frac{1}{6}\, \mathcal{P}_\Sigma(z,\bar{z})\, + \, \frac{1}{6} \, \mathfrak{f}_{I} \, \mathfrak{C}_\Sigma^I \right) \, D^\Sigma \, + \, \frac{1}{6} M^\Lambda \, M^\Sigma \, k^i_\Lambda \, k_\Sigma^{j^\star} \, g_{ij^\star} \nonumber\\
&& +\,\frac{1}{6} \left(\mathcal{H}^i \, \partial_i W \, + \,
\mathcal{H}^{\ell^\star} \,
\partial_{\ell^\star} {\overline{W}} \right) \, - \, \frac{1}{6} \, g_{i\ell^\star} \,\mathcal{H}^i \,
\mathcal{H}^{\ell^\star} \, - \, \frac{1}{2} \,\mathbf{e}\,
\mathbf{\kappa}_{\Lambda\Sigma}\, D^\Lambda \, D^\Sigma
\end{eqnarray}
In the above equation the vectors $\mathfrak{C}_\Sigma^I$ ($I\, = \,
1, \dots, r$) project onto the $r$ independent generators of the
center of the gauge Lie algebra $Z(\mathbb{G})$. For each of these
generators one can add a separately supersymmetric invariant term,
named Fayet Iliopoulos term \cite{Fayet:1974jb}, which is just
linear in the corresponding auxiliary fields $D_I \, \equiv \,
\mathfrak{C}_\Sigma^I \, D^\Sigma$. Namely we have:
\begin{equation}\label{FItermo}
  \mbox{Fayet Iliopolus term} \, \equiv \,  \mathfrak{f}_{I} \, \mathfrak{C}_\Sigma^I  \, D^\Sigma  \, \epsilon_{mnp} \, e^m \, \wedge \, e^n \, \wedge \, e^p
\end{equation}
where $\mathfrak{f}_I$ are independent constants (the Fayet
Iliopoulos constants). Furthermore:
\begin{equation}\label{Psigma}
    \mathcal{P}_{\Sigma}(z,\bar{z}) \, = \, - \, {\rm i} \, \left( k^i_\Lambda \, \partial_i \, \mathcal{K} \, - \, k^{i^\star}_\Lambda \, \partial_{i^\star} \, \mathcal{K}\right)
\end{equation}
is the moment map of the gauged holomorphic Killing vectors
satisfying the identity:
\begin{equation}\label{killogenero}
    k^i_\Sigma \, = \, {\rm i} \, g^{ij^\star} \, \partial_{j^\star}\, \mathcal{P}_{\Sigma} \quad ; \quad k^{i^\star}_\Sigma \, = \, -  \, {\rm i} \, g^{i^\star j} \, \partial_j\, \mathcal{P}_{\Sigma}
\end{equation}
\subsection{Structure of the scalar potential and of the Lagrangian
after the elimination of the auxiliary fields} Let us first observe
that the structure of the theory is substantially different when the
standard kinetic term of the gauge fields are included, namely when
$\mathbf{e} \ne 0$, and when they are not $\mathbf{e} = 0$. Hence we
discuss the two cases separately.
\subsubsection{$\mathcal{N}=2$ Pure Chern Simons Gauge Theory} When
$\mathbf{e} = 0$, the lagranian takes the following form:
\begin{eqnarray}
\mathcal{L}_{CSoff}&=& - \, \alpha \, \mbox{Tr} \,\left(\mathfrak{F}
\, \wedge \, \mathcal{A} \, + \, \frac{2}{3}\, \mathcal{A}\wedge \,
\mathcal{A}\wedge \, \mathcal{A}\right)\,+\,\left( \frac{1}{2} \,
g_{ij^\star} \, \Pi^{m\mid i} \, \nabla \bar{z}^{j^\star} \, + \,
\bar{\Pi}^{m\mid j^\star} \, \nabla z^{i}\right)\, \wedge \, e^n \,
\wedge \,  e^p \, \epsilon_{mnp}\nonumber\\
\null &&  - \, \frac{1}{6} \, \, g_{ij^\star} \, \Pi^{m\mid i} \,
\bar{\Pi}^{m\mid j^\star} \, e^r \,  \wedge \, e^s \, \wedge \,
e^t \, \epsilon_{rst}\nonumber\\
&& +\,{\rm i}\frac{1}{2} \, g_{ij^\star} \,
\left(\bar{\chi}^{j^\star} \, \gamma^m \,\nabla \chi{i} \, + \,
\bar{\chi}^{i}_c \, \, \gamma^m \,
\nabla \chi{i^\star}_c\right) \, \wedge \, e^n \, \wedge \,  e^p \, \epsilon_{mnp} \nonumber\\
&& \left( \, - \, \frac{1}{3} \, M^\Lambda \,\left( \partial_i
k^j_\Lambda \, g_{j\ell^\star} \, \bar{\chi}^{\ell^\star}
\,\chi^i \,+ \,  \partial_{i^\star} k^{j^\star}_\Lambda \,
g_{j\ell^\star} \bar{\chi}^\ell_c \, \chi^{i^\star}_c \right) \, +
\, \frac{\alpha}{3} \, \left( \bar{\lambda}^\Lambda \,
\lambda^\Sigma \, + \, \bar{\lambda}^\Lambda_c \,
\lambda^\Sigma_c\right) \,
\mathbf{\kappa}_{\Lambda\Sigma}\right.\nonumber\\
&&\left. \, +\, {\rm i}\, \frac{1}{3} \, \left(
\bar{\chi}^{j^\star}_c \, \lambda^\Lambda \, k^i_\Lambda \, - \,
\bar{\chi}^{i}_c \, \lambda^\Lambda \, k^{j^\star}_\Lambda\right) \,
g_{ij^\star} \, \right.\nonumber\\
&&\left. + \, \frac{1}{6} \, \left( \partial_i\partial_j W\,
\bar{\chi}^i_c \, \chi^j \, + \,
\partial_{i^\star}\partial_{j^\star} {\overline{W}}\,
\bar{\chi}^{i^\star} \, \chi^{j^\star}_c\right) \right)\, \wedge \,
e^n \, \wedge \,  e^p \, \epsilon_{mnp} \nonumber\\
&&-V\left(M,D,\mathcal{H},z,\bar{z}\right)  \, \epsilon_{mnp} \, e^m
\, \wedge \, e^n \, \wedge \, e^p \label{pastiglialeone}
\end{eqnarray}
where the potential in terms of physical and auxiliary fields is the following one:
\begin{eqnarray}
V\left(M,D,\mathcal{H},z,\bar{z}\right)&=& \left(\frac{\alpha}{3} \,
M^\Lambda \,\mathbf{\kappa}_{\Lambda\Sigma}\,  - \, \frac{1}{6}\,
\mathcal{P}_\Sigma(z,\bar{z})\, + \, \frac{1}{6} \, \mathfrak{f}_{I}
\, \mathfrak{C}_\Sigma^I \right) \, D^\Sigma \, + \, \frac{1}{6}
M^\Lambda \, M^\Sigma \, k^i_\Lambda \, k_\Sigma^{j^\star} \,
g_{ij^\star} \nonumber\\
&& +\,\frac{1}{6} \left(\mathcal{H}^i \, \partial_i W \, + \,
\mathcal{H}^{\ell^\star} \,
\partial_{\ell^\star} {\overline{W}} \right) \, - \, \frac{1}{6} \, g_{i\ell^\star} \,\mathcal{H}^i \,
\mathcal{H}^{\ell^\star} \label{karamella}
\end{eqnarray}
In this case the gauge multiplet does not propagate and it is
essentially made of lagrangian multipliers for certain constraints.
Indeed  the auxiliary fields,  the gauginos and the vector multiplet
scalars have algebraic field equations so that they can be
eliminated through the solutions of such equations of motion. The
vector multiplet auxiliary scalars $D^\Lambda$ appear only as
lagrangian multipliers of the constraint:
\begin{equation}\label{Msolvo}
    M^\Lambda \, = \, \frac{1}{2\alpha} \, {\mathbf{\kappa}}^{\Lambda \Sigma}\,\left(   \mathcal{P}_\Sigma \, -\, \mathfrak{f}_I \,  \mathfrak{C}_\Sigma^I \right)
\end{equation}
while the variation of the auxiliary fields $\mathcal{H}^{j^\star}$
of the Wess Zumino multiplets yields:
\begin{equation}\label{eliminoH}
\mathcal{H}^{i} \, = \, g^{ij^\star} \, \partial_{j^\star} \,
\overline{W} \quad ; \quad \overline{\mathcal{H}}^{j^\star} \, = \,
g^{ij^\star} \, \partial_{i} \, {W}
\end{equation}
On the other hand, the equation of motion of the field $M^\Lambda$
implies:
\begin{equation}\label{gospadi}
D^\Lambda \, = \, - \, \frac{1}{\alpha} \, \mathbf{\kappa}^{\Lambda
\Gamma} g_{ij^\star} \, k_\Gamma^{i} \, k^{j^\star}_\Sigma\,
M^\Sigma \, = \, - \, \frac{1}{2\,\alpha^2} \,g_{ij^\star}\,
\mathbf{\kappa}^{\Lambda \Gamma}\, k_\Gamma^{i} \,
k^{j^\star}_\Sigma\,\mathbf{\kappa}^{\Sigma \Delta}\,\left(
\mathcal{P}_\Delta \, -\, \mathfrak{f}_I \,  \mathfrak{C}_\Delta^I
\right)
\end{equation}
which finally resolves all the auxiliary fields in terms of
functions of the physical scalars.
\par
Upon use of both constraints (\ref{Msolvo}) and (\ref{eliminoH}) the
scalar potential takes the following positive definite form:
\begin{eqnarray}
    V(z,\bar{z}) & = & \frac{1}{6} \,
\left( \partial_i W \, \partial_{j^\star} \overline{W}
\,g^{ij^\star} \, + \, \mathbf{m}^{\Lambda\Sigma} \,
\left(\mathcal{P}_\Lambda \, - \, \mathfrak{f}_I \,
\mathfrak{C}_\Lambda^I\right) \,
\left(\mathcal{P}_\Sigma \, - \, \mathfrak{f}_I \, \mathfrak{C}_\Sigma^I \right) \right)\nonumber\\
\mathbf{m}^{\Lambda\Sigma}(z,\bar{z}) & \equiv &
\frac{1}{4\alpha^2}\,\mathbf{\kappa}^{\Lambda\Gamma} \,
\mathbf{\kappa}^{\Sigma\Delta} \, k_\Gamma^i \, k_\Delta^{j^\star}
\, g_{ij^\star} \label{quadraPot}
\end{eqnarray}
In a similar way the gauginos can be resolved in terms of the
chiralinos:
\begin{equation}\label{eliminogaugino}
\lambda^\Lambda \, = \, -\, \frac{1}{2\alpha} \,
\mathbf{\kappa}^{\Lambda\Sigma} \, g_{ij^\star} \chi^i \,
k^{j^\star}_\Sigma \quad ; \quad \lambda^\Lambda_c \, = \, -\,
\frac{1}{2\alpha} \, \mathbf{\kappa}^{\Lambda\Sigma} \, g_{ij^\star}
\chi^{j^\star} \, k^{i}_\Sigma
\end{equation}
In this way if we were able to eliminate also the gauge one form $\mathcal{A}$ the Chern Simons gauge theory
would reduce to a theory of Wess-Zumino multiplets with additional interactions. The elimination of
$\mathcal{A}$, however, is not possible in the non-abelian case and it is possible in the abelian case only
through duality non local transformations.  This is the corner where interesting non perturbative dynamics is
hidden.
\subsubsection{$\mathcal{N}=2$ Maxwell-Chern-Simons Gauge Theories}
Of interest are also the mixed Maxwell-Chern-Simons Gauge Theories
where both the Maxwell and Chern Simons kinetic terms are included,
namely where $\mathbf{e}\ne0$ and $\alpha \ne 0$. In this case the
gauge fields propagate and so do the gauginos and the vector
multiplet scalars $M^\Lambda$.
\par
At the level of the potential the main difference is the presence of
the quadratic term in the vector multiplet auxiliary fields
$D^\Lambda$. Eliminating these latter through their own field
equations and similarly doing with the auxiliary fields of the WZ
multiplets we get the following potential for the propagating
scalars $M^\Lambda$ and $z^i$:
\begin{eqnarray}\label{MCSpotus}
V_{MCS} \left(M,z,\bar{z}\right)& =&\frac{1}{2\, {\mathbf{e}}^2}\,
\mathbf{\kappa}^{\Lambda\Sigma}\, \left(\frac{\alpha}{3} \,
M^\Delta \,\mathbf{\kappa}_{\Lambda\Delta}\,  - \, \frac{1}{6}\,
\mathcal{P}_\Lambda(z,\bar{z})\, + \, \frac{1}{6} \,
\mathfrak{f}_{I} \, \mathfrak{C}_\Lambda^I \right) \, \times \,
\nonumber\\
&& \left(\frac{\alpha}{3} \,  M^\Gamma
\,\mathbf{\kappa}_{\Sigma\Gamma}\,  - \, \frac{1}{6}\,
\mathcal{P}_\Sigma(z,\bar{z})\, + \, \frac{1}{6} \, \mathfrak{f}_{I}
\, \mathfrak{C}_\Sigma^I \right)\, \nonumber\\
&&+ \, \frac{1}{6}\left( \mathbf{m}_{\Lambda\Sigma}(z,\bar{z})
M^\Lambda \, M^\Sigma \, + \, g^{ij^\star} \partial_i W \,
\partial_{j^\star} \overline{W}\right)
\end{eqnarray}
Let us now consider the other terms in the Lagrangian and perform
the transition from the first to the second order formalism by
eliminating the remaining auxiliary fields. The final form of the
second order Lagrangian for the most general Maxwell-Chern-Simons
matter coupled gauge-theory is the following one:
\begin{eqnarray}
\mathcal{L}_{MCS}^{gen} &=& - \, \frac{1}{6}\,
\mathbf{\kappa}_{\Lambda\Sigma}\,\left[\alpha \,
\left(F^\Lambda_{mn} A^\Sigma_p \, +\, \frac{2}{3}\,
f_{\phantom{\sigma}\Gamma\Delta}^\Sigma A^\Lambda_m\,  A^\Gamma_n\,
A^\Delta_p\right) \, \epsilon^{mnp}
\, + \, \mathbf{e} \, F^\Lambda_{mn} \, F^\Sigma_{mn} \right] \nonumber\\
&& + \, \mathbf{e} \, \frac{1}{12} \,
\mathbf{\kappa}_{\Lambda\Sigma}\, \nabla_m M^\Lambda \, \nabla_m
M^\Sigma \, + \, \frac{1}{6} \, g_{ij^\star} \, \nabla_m z^i  \,
\nabla_m z^{j^\star} \nonumber\\
&& -\,{\rm i}\, \mathbf{e}\, \frac{1}{12} \, \left(\bar{
\lambda}^\Lambda \, \gamma^m \, \nabla_m\lambda^\Lambda \, + \,
\bar{ \lambda}^\Lambda_c \, \gamma^m \, \nabla_m\lambda^\Lambda_c
\right) \, + \, {\rm i}\, \frac{1}{12} \, g_{ij^\star} \left(\bar{
\chi}^{j^\star} \, \gamma^m \, \nabla_m\chi^i \, + \, \bar{ \chi}^i
\, \gamma^m \, \nabla_m\chi^{j^\star} \right) \nonumber\\
&& \left. \, - \, \frac{1}{3} \, M^\Lambda \,\left( \partial_i
k^j_\Lambda \, g_{j\ell^\star} \, \bar{\chi}^{\ell^\star}
\,\chi^i \,+ \,  \partial_{i^\star} k^{j^\star}_\Lambda \,
g_{j\ell^\star} \bar{\chi}^\ell \, \chi^{i^\star} \right) \, + \,
\frac{\alpha}{3} \, \left( \bar{\lambda}^\Lambda \, \lambda^\Sigma
\, + \, \bar{\lambda}^\Lambda_c \, \lambda^\Sigma_c\right) \,
\mathbf{\kappa}_{\Lambda\Sigma}\right.\nonumber\\
&&\left. \,+\,{\rm i} \, \frac{1}{3}\, \left( \bar{\chi}^{j^\star}
\, \lambda^\Lambda \, k^i_\Lambda \, - \, \bar{\chi}^{i} \,
\lambda^\Lambda \, k^{j^\star}_\Lambda\right) \, g_{ij^\star} \, -
\, \mathbf{e} \,\frac{1}{6} \,\mbox{Tr} \left(M \,
\left(\left[\lambda\, , \, \lambda\right] \, + \, \left[\lambda_c\,
, \, \lambda_c\right] \right)\right)\right.\nonumber\\
&&\left.+ \, \frac{1}{6} \, \left( \partial_i\partial_j W\,
\bar{\chi}^i \, \chi^j \, + \, \partial_{i^\star}\partial_{j^\star}
{\overline{W}}\, \bar{\chi}^{i^\star} \, \chi^{j^\star}\right) \, -
\, V_{MCS} \left(M,z,\bar{z}\right) \right. \label{generalagra}
\end{eqnarray}
We consider next special instances of theories inside the above
described general families.
\subsection{${\cal N}=3$ Chern Simons
gauge theory in three dimensions}\label{N=3gaugetheory} In this
section we discuss the structure of a three dimensional Chern Simons
gauge theory with ${\cal N}=3$ supersymmetry. The starting point is
the discussion of a complete $\mathcal{N}=3$ gauge theory in the
same dimensions. The ${\cal N}=3$ case is just a particular case in
the  class of theories described in the previous section since a
theory with ${\cal N}=3$ SUSY, must {\it a fortiori} be an ${\cal
N}=2$ theory. In \cite{Fabbri:1999ay} , the case of ${\cal N}=4$
theories was also considered, within the ${\cal N}=2$ class. These
latter are obtained through dimensional reduction of an ${\cal
N}_4=2$ theory in four--dimensions. Indeed since each $D=4$ Majorana
spinor splits, under dimensional reduction on a circle
$\mathbb{S}^1$, into two $D=3$ Majorana spinors, the number of
three--dimensional supercharges is just twice the number of $D=4$
supercharges:
\begin{equation}
  {\cal N}_3 = 2 \, \times \, {\cal N}_4
\label{n34}
\end{equation}
The ${\cal N}_3=3$ case corresponds to an intermediate situation. It
is an ${\cal N}_3=2$ theory with the field content of an ${\cal
N}_3=4$ one, but with additional ${\cal N}_3=2$ interactions that
respect  three out of the four supercharges obtained through
dimensional reduction. Using an ${\cal N}=2$ superfield formalism
and the notion of twisted chiral multiplets it was shown in
\cite{kapustin} that for abelian gauge theories these additional
${\cal N}_3=3$ interactions are
\begin{enumerate}
  \item A Chern Simons term, with coefficient $\alpha$
  \item A mass-term  with coefficient $\mu=\alpha$ for the
  chiral field $Y^I$ in the adjoint of the color gauge group.
  By this latter we denote the complex field belonging,
  in four dimensions, to the ${\cal N}_4=2$ gauge
  vector multiplet.
\end{enumerate}
In \cite{ringoni} the authors  retrieved for non-abelian gauge
theories the same result as that found by the authors of
\cite{kapustin} for the abelian theories. In \cite{ringoni} the
construction was presented  in the component formalism which is
better suited to discus the relation between the world--volume gauge
theory and the geometry of the transverse cone
$\mathcal{C}(\mathcal{M}_7)$. Let us also remark that the arguments
used in\cite{Aharony:2008ug}  are the same which were spelled out
ten years earlier in \cite{ringoni}. In this section we summarize in
the more general notations based on HyperK\"ahler metric and the
triholomorphic moment maps the general form of a non abelian $ {\cal
N}=3$ Chern Simons gauge theory in three dimensions as it was
obtained in \cite{ringoni}.
\subsubsection{The field content and the interactions} The strategy
of \cite{ringoni} was that of writing the ${\cal N}=3$ gauge theory
as a special case of an ${\cal N}=2$ theory, whose general form was
discussed in previous sections. For this latter the field content is
given by:
\begin{equation}
   \begin{array}{|c|c|c|c|}
   \hline
     \mbox{multipl. type $\, /\,\mathrm{SO(1,2)}$ spin}  &  1 & \ft 1 2 & 0 \\
     \hline
     \hline
     \null & \null &\null & \null \\
     \mbox{vector multipl.} &  \underbrace{A^I_\mu}_{\mbox{gauge field}} &
     \underbrace{\left( \lambda^{+I},\lambda^{-I}
     \right)}_{\mbox{gauginos}}  &\underbrace{ M^I}_{\mbox{real scalar}} \\
     \hline
     \null &  \null &\null & \null \\
     \mbox{chiral multip.} &  \null &\underbrace{\left( \chi^{+i},\chi^{-i^*}
     \right)}_{\mbox{chiralinos}}  & \underbrace{ z^i,\ \bar z^{i^*}}_
     {\mbox{complex scalars}}
     \\
     \hline
   \end{array}
\label{fieldcont}
\end{equation}
and  the complete Lagrangian was given in the previous sections. In
particular the complete Chern Simons Lagrangian  before the
elimination of the auxiliary fields was displayed in
eq.(\ref{pastiglialeone}).
\par
The Chern Simons ${\cal N}=3$ case is obtained when the  following
conditions are fulfilled:
\begin{itemize}
\item The spectrum of chiral multiplets is made of $\mbox{dim}
\mathcal{G} \, + \, 2n$ complex fields arranged in the following way
 \begin{equation}\label{cagullo}
   z^i \, = \, \left\{ \begin{array}{rccl}
   Y^\Lambda & = & \mbox{complex fields} & \mbox{in the \textbf{adjoint rep. of the color group}} \\
   q^\alpha & = & \left( \begin{array}{c} u^a \\ v_b \\\end{array}\right) & \left \{
   \begin{array}{l}
   \mbox{$2n$ complex fields spanning a \textbf{HyperK\"ahler manifold} $\mathrm{HK}_{2n}$}\\
   \mbox{which is invariant under a }\\
   \mbox{\textbf{triholomorphic action} of the gauge group $\mathcal{G}$}\\
   \end{array} \right.\\
   \end{array} \right.
.\end{equation}
\item the K\"ahler potential has the following form:
\begin{equation}\label{kallettus}
  \mathcal{K}(Y,u,v) \, = \, \hat{\mathcal{K}}(u,v)
\end{equation}
where $\hat{\mathcal{K}}(u,v)$ is the K\"ahler potential of the
Ricci-flat HyperK\"ahler metric of the HyperK\"ahler manifold
$\mathrm{HK}_{2n}$. The assumption that $ \mathcal{K}(Y,u,v)$ does
not depend on $Y^\Lambda$ implies that the kinetic term of  these
scalars vanishes turning them into auxiliary fields that can be
integrated away.
  \item The superpotential $W(z)$ has the following form:
  \begin{equation}
  W(Y,u,v)=\kappa_{\Lambda\Sigma}\,\left(Y^I\, \mathcal{P}_+^\Sigma(u,v)\,
  +\,2\,\alpha\,\,Y^\Lambda\,Y^\Sigma\right)
  \label{suppotn3}
  \end{equation}
where $\mathcal{P}^\Sigma_+(u,v)$ denotes the holomorphic part of
the triholomorphic moment map induced by the triholomorphic action
of the color  group on $\mathrm{HK}_{2n}$.
\end{itemize}
The reason why these two choices make the theory ${\cal N}_3=3$
invariant is simple: the first choice corresponds to assuming the
field content of an ${\cal N}_3=4$ theory which is necessary since
${\cal N}_3=3$ and ${\cal N}_3=4$ supermultiplets are identical. The
second choice takes into account that the metric of the
hypermultiplets must be HyperK\"ahler and that the gauge coupling
constant was sent to infinity. The third choice introduces an
interaction that preserves ${\cal N}_3=3$ supersymmetry but breaks
(when $\alpha \ne 0$) ${\cal N}_3=4$ supersymmetry.
\par
Going back to the off-shell Chern Simons lagrangian given in
eq.(\ref{pastiglialeone}) one can perform the elimination of the
auxiliary fields that now include $Y^\Lambda,D^\Lambda,M^\Lambda,
\mathcal{H}^i$ at the bosonic level and the gauginos
$\lambda^\Lambda, \lambda^\Lambda_c,\chi^\Lambda,\chi^\Lambda_c$ at
the fermionic level (Note that there are two more non propagating
gauginos coming from the chiral multiplet in the adjoint
representation of the gauge group). We do not enter the details of
the integration over the non propagating fermions and we just
consider the bosonic lagrangian emerging from the integration over
the auxiliary bosonic fields. The first integration to perform is
that over the auxiliary field $\mathcal{H}^\Lambda$. This is simply
the lagrangian multiplier of the constraint:
\begin{equation}\label{cartacallus}
  \partial_\Lambda W \, = \, 0 \quad \Rightarrow \quad Y^\Lambda \, = \, \frac{1}{4\alpha} \, \mathcal{P}^\Lambda_+(u,v)
\end{equation}
Substituting this back into the lagrangian  yields a potential with
the same structure as that in eq.(\ref{quadraPot}) but with a
modified superpotential which becomes quadratic in the holomoprhic
momentum map:
\begin{eqnarray}
    V(u,v) & = & \frac{1}{6} \,
\left( \partial_\alpha \mathfrak{W} \, \partial_{\beta^\star}
\overline{\mathfrak{W}} \,g^{\alpha\beta^\star} \, + \,
\mathbf{m}^{\Lambda\Sigma} \,
\mathcal{P}^3_\Lambda \, \mathcal{P}^3_\Sigma \, \right) \nonumber\\
\mathbf{m}^{\Lambda\Sigma}(u,v) & \equiv &
\frac{1}{4\alpha^2}\,\mathbf{\kappa}^{\Lambda\Gamma} \,
\mathbf{\kappa}^{\Sigma\Delta} \, k_\Gamma^\alpha \, k_\Delta^{\beta^\star} \, g_{\alpha\beta^\star} \\
\mathfrak{W}& = & - \frac{1}{8\alpha} \, \mathcal{P}_+^\Lambda \,
\mathcal{P}_+^\Sigma \, \kappa_{\Lambda\Sigma} \label{quartaPot}
\end{eqnarray}
\subsubsection{The ${\cal N}=3$ gauge theory corresponding to the
$\n010$ compactification}\label{theoryofN010} Having clarified the
structure of a generic ${\cal N}=3$ gauge theory let us consider, as
an illustration, the specific one associated with the $\n010$
seven--manifold following the presentation of \cite{ringoni}. As
explained in \cite{Fabbri:1999hw} (see eq.(B.1) of that paper) the
manifold $\n010$ is the circle bundle inside $\mathcal{O}(1,1)$ over
the flag manifold $\mathbb{F}(1,2;3)$. In other words we have
\begin{equation}
  \n010 \, \stackrel{\pi}{\longrightarrow}\mathbb{F}(1,2;3)
\label{fibrsuflag}
\end{equation}
where, by definition,
\begin{equation}
  \mathbb{F}(1,2;3) \equiv \frac{\mathrm{SU(3)}}{H_1 \times H_2}
\label{flagga}
\end{equation}
is  the homogeneous space obtained by modding $\mathrm{SU(3)}$ with
respect to its maximal torus:
\begin{equation}
  H_1 = \exp \left[i \theta_1 \left( \begin{array}{ccc}
    1 & 0 & 0 \\
    0 & -1 & 0 \\
    0 & 0 & 0 \
  \end{array}\right) \right] \quad ; \quad
   H_2 = \exp \left[i \theta_2 \left(\begin{array}{ccc}
     1 & 0 & 0 \\
     0 & 1 & 0 \\
     0 & 0 & -2 \\
   \end{array} \right) \right]
\label{maxtor}
\end{equation}
Furthermore as also explained in \cite{Fabbri:1999hw} (see
eq.(B.2)), the base manifold $\mathbb{F}(1,2;3)$ can be
algebraically described as the following quadric
\begin{equation}
  \sum_{i=1}^{3} \, u^i \, v_i = 0
\label{vanlocus}
\end{equation}
in $\mathbb{P}^2 \times \mathbb{P}^{2*}$, where $u^i$ and $v_i$ are
the homogeneous coordinates of $\mathbb{P}^2$ and $\mathbb{P}^{2*}$,
respectively.
\par
Hence a complete description of the metric cone $\mathcal{C}\left(
\n010\right) $ can be given by writing the following equations in $
\mathbb{C}^3 \times \mathbb{C}^{3*}$:
\begin{equation}
 \mathcal{C}\left(
 \n010\right)= \left \{ \begin{array}{rcll}
    |u^i|^2-|v_i|^2 & = & 0 & \mbox{fixes equal the radii of $\mathbb{P}^2$
    and $\mathbb{P}^{2*}$}  \\
    2 \, u^i \, v_i & = & 0 & \mbox{cuts out the quadric locus} \\
    \left( u^i \, e^{i\theta} , v_i \, e^{-i\theta}\right)
    &\simeq &\left( u^i,v_i\right) &\mbox{identifies points of $\mathrm{U(1)}$
    orbits}\
  \end{array}\right.
\label{metrcon}
\end{equation}
Eq.s (\ref{metrcon}) can be easily interpreted as the statement that
the cone ${\cal C}\left(\n010\right)$ is the HyperK\"ahler quotient
of a flat three-dimensional quaternionic space with respect to the
triholomorphic action of a $\mathrm{U(1)}$ group. Indeed the first
two equations in (\ref{metrcon}) can be rewritten as the vanishing
of the triholomorphic moment map of a $\mathrm{U(1)}$ group. It
suffices to identify:
\begin{eqnarray}
{\cal P}_3 & = & -\left( |u^i|^2 -|v_i|^2 \right)  \nonumber\\
{\cal P}_- & =& 2 v_i u^i \label{agniu1}
\end{eqnarray}
Comparing with eq.s (\ref{quartaPot}) we see that the cone
$\mathcal{C} (\n010)$ can be correctly interpreted as the space of
classical vacua in an abelian ${\cal N}=3$ gauge theory with $3$
hypermultiplets in the fundamental representation of a flavor group
$\mathrm{SU(3)}$. Indeed if the color group is $\mathrm{U(1)}$ there
is only one value for the index $\Lambda$. The potential is a
positive definite quadratic form in the moment maps with minimum at
zero which is attained when the moment map vanishes.
\par
Relying on this geometrical picture of the transverse space to an
$M=2$--brane leaving on AdS$_4 \times \n010$, in \cite{ringoni} it
was conjectured that the ${\cal N}=3$ non--abelian gauge theory
whose infrared conformal point is dual to $D=11$ supergravity
compactified on AdS$_4 \times \n010$ should have the following
structure:
\begin{equation}
 \begin{array}{crcl}
   \mbox{gauge group} & \mathcal{ G}_{gauge} & = & \mathrm{SU(N)}_1 \times
   \mathrm{SU(N)}_2 \\
   \null & \null & \null & \null \\
   \mbox{flavor group} & \mathcal{G}_{flavor} & = & \mathrm{SU(3)}  \\
   \null & \null & \null & \null \\
   \mbox{color representations of the hypermultiplets} & \left[ \begin{array}{c}
  u  \\
  v
\end{array} \right] & \Rightarrow & \left[ \begin{array}{c}
  \left({\bf N}_1,{\bf \bar N}_2\right)  \\
  \left({\bf \bar N}_1, {\bf N}_2\right)
\end{array} \right]  \\
\null & \null & \null & \null \\
\mbox{flavor representations of the hypermultiplets} & \left[
\begin{array}{c}
  u  \\
  v
\end{array} \right] & \Rightarrow & \left[ \begin{array}{c}
  {\bf 3}  \\
 {\bf \bar 3}
\end{array} \right]  \\
 \end{array}
\label{assegnati}
\end{equation}
More explicitly and using an ${\cal N}=2$ notation we can say that
the field content of the theory proposed in \cite{ringoni} is given
by the following chiral fields, that are all written as $N \times N$
matrices:
\begin{equation}
  \begin{array}{cccc}
    Y_1 & = & \left(Y_1\right)^{\Lambda_1}_{\phantom{\Lambda_1}\Sigma_1 } &
    \mbox{adjoint of $\mathrm{SU(N)}_1$} \\
   Y_2 & = & \left(Y_2\right)^{\Lambda_2}_{\phantom{\Lambda_2}\Sigma_2}  &
    \mbox{adjoint of $\mathrm{SU(N)}_2$ }\\
     u^i & = & \left(u^i\right)^{\Lambda_1}_{\phantom{\Lambda_1}\Sigma_2 } &
    \mbox{in the $({\bf 3},{\bf N}_1,{\bf \bar N}_2)$} \\
    v_i & = & \left(v_i\right)^{\phantom{\Sigma_1}\Lambda_2}_{\Sigma_1}  &
    \mbox{in the $({\bf 3},{\bf \bar N}_1,{\bf N}_2)$} \\
  \end{array}
\label{matricicole}
\end{equation}
and the superpotential before integration on the auxiliary fields
$Y$ can be written as follows:
\begin{equation}
  W = 2 \, \left [\mbox{Tr} \left( Y_1\, u^i \, v_i \right) +
  \mbox{Tr}\left( Y_2\, v_i \, u_i \right) +  \alpha_1 \mbox{Tr}\left(
  Y_1 \, Y_1 \right ) + \alpha_2 \, \mbox{Tr} \left( Y_2 \, Y_2 \right) \right]
\label{suppotmat}
\end{equation}
where $\alpha_{1,2}$ are the  Chern Simons coefficients associated
with the $\mathrm{SU(N)}_{1,2}$ simple gauge groups, respectively.
Setting:
\begin{eqnarray}
\alpha_1 & = & \pm \alpha_2 = \alpha \label{galpha}
\end{eqnarray}
and integrating out the two fields $Y_{1,2}$ that have received a
mass by the Chern Simons mechanism in \cite{ringoni} it was obtained
the following effective quartic superpotential:
\begin{equation}
  W^{eff}= -\ft 1 2 \, \ft {1}{\alpha} \left[
  \mbox{Tr} \left( v_i \, u^i \, v_j \, u^j  \right) \pm
  \mbox{Tr} \left( u^i \, v_i \, u^j \, u_j\right)   \right]
\label{effepot}
\end{equation}
The vanishing relations one can derive from the above superpotential
are the following ones:
\begin{equation}
  u^i \, v_j \, u^j = \pm u^j \, v_j \,  u^i \quad ; \quad
   v_i \, u^j \, v_j = \pm v_j \, u^j \,  v_i
\label{vanrel}
\end{equation}
Consider now the chiral conformal superfields one can write in this
theory:
\begin{equation}
  \Phi^{i_1\, i_2 \, \dots \, i_k}_{j_1 \, j_2 \, \dots \, j_k}
  \equiv
  \mbox{Tr} \left( u^{(i_1} \, v_{(j_1} \, u^{i_2} \, v_{j_2} \, \dots
  \, u^{i)_k} \, v_{j_k)} \right)
\label{chiralop}
\end{equation}
where the round brackets denote symmetrization on the indices. The
above operators have $k$ indices in the fundamental representation
of $\mathrm{SU(3)}$ and $k$ indices in the antifundamental one, but
they are not yet assigned to the irreducible representation:
\begin{equation}
  M_1=M_2 = k
\label{irredu}
\end{equation}
as it is predicted both by general geometric arguments and by the
explicit evaluation of the Kaluza Klein spectrum of hypermultiplets
\cite{Fre':1999xp}. To be irreducible the operators (\ref{chiralop})
have to be traceless. This is what is implied by the vanishing
relation (\ref{vanrel}) if we choose the minus sign in
eq.(\ref{galpha}).
\par
In \cite{ringoni} it was noticed that for $N^{0,1,0}$ the form of
the superpotential, which is dictated by the Chern-Simons term, is
strongly reminiscent of the superpotential considered in
\cite{witkleb}. Indeed the CFT theory associated with $N^{0,1,0}$
has  many analogies with the simpler cousin $T^{1,1}$
\cite{sergiotorino}. However it was stressed in \cite{ringoni} that
there is also a crucial difference, pertaining to a general
phenomenon that was discussed for the case of compactifications on
$M^{1,1,1}$ and $Q^{1,1,1}$ in \cite{Fabbri:1999mk} and
\cite{Fabbri:1999hw}. The moduli space of vacua of the abelian
theory is isomorphic to the cone ${\cal C}\left( \n010\right)$. When
the theory is promoted to a non-abelian one, there are naively
conformal operators whose existence is in contradiction with
geometric expectations and with the KK spectrum, in this case the
hypermultiplets that do not satisfy relation~(\ref{irredu}).
Differently from what happens for $T^{1,1}$ \cite{witkleb}, the
superpotential in eq. (\ref{effepot}) is not sufficient for
eliminating these redundant non-abelian operators.
\par
Ten years later in a paper by Gaiotto et al \cite{Gaiotto:2009tk},
it was advocated that, maintaining the same flavor-group assignments
and the same color group, the  color representation assignments of
the hypermultiplets that lead to the correct dual CFT are slightly
different from those shown in eq. (\ref{matricicole}) since in
addition to the bi-fundmental representation one needs also the two
fundamental ones.
\par
In any case it is appropriate to stress that, on the basis of the
general form of the $\mathcal{N}=3$ gauge theory discussed above as
a particular case of the general $\mathcal{N}=2$ theory, it was just
in \cite{ringoni} that the structure of an $\mathcal{N}=3, D=3$
Chern-Simons gauge theory, corner stone of the famous ABJM
model\cite{Aharony:2008ug}, was for the first time derived in the
literature. Indeed in \cite{ringoni} it was just conjectured that
the gauge coupling constant flows to infinity at the infrared
conformal point, so that the effective lagrangian is obtained from
the general one by letting $\mathbf{e} \to 0$. It was in
\cite{ringoni} that the conversion of the  $Y^\Lambda$ field into a
lagrangian multiplier was for the first time observed, leading to
the generation of an effective superpotential of type
(\ref{effepot}).
\par
In this paper we continue to explore the properties of the general $\mathcal{N}=2$, $D=3$ gauge theory both
from the point of view of its formal structure in superspace and as a starting point for the dual gauge
theories associated with M--theory probing $\mathbb{C}^n/\Gamma$ singularities and their resolutions. The
mathematical aspects of $\mathbb{C}^3/\Gamma$ resolutions in relation with the construction of $D=3$ Chern
Simons gauge theories is the topic of a forthcoming paper by one of us in collaboration with Ugo Bruzzo that
is currently in progress \cite{ugopietro}.
\section{Integral forms in superspace and three--dimensional
Chern--Simons gauge theories}
In the present section, we reconsider the construction of the action of the Chern Simons gauge theories under
investigation by using the method of \textit{Integral Forms} and of \textit{Picture Changing Operators}
(PCO's) developed in \cite{Castellani:2015paa,Catenacci:2016qzd,Grassi:2016apf,Castellani:2016ibp}. For that
purpose, we briefly describe the principles of this method and we give some of the relevant results without a
complete derivation. The latter will be published elsewhere \cite{N=2_CS_pietro} since the details of those
derivations are not important for the scope of the present paper.
\par
The rhenomic action ${\cal L}^{{\cal N}=2}_{rheo}$, decomposed into pieces as in eqs. (\ref{MM1}--\ref{MM5}),
is our starting point. It is a 3-form on the superspace ${\cal M}^{(3|4)}$ parametrized locally by the
variables $(x^m, \theta, \theta_c)$ with dimensions $(3|4)$. The superspace is described in sec.
\ref{supergeo} and the supervielbeins $(e^m, \psi, \psi_c)$ form a supervector. The supervielbein, expanded
on the anholonomic basis, can be represented by a supermatrix $(3|4) \times (3|4)$. The rigid superspace is
flat, but it has torsion.
\par
The geometrical approach to the supersymmetric field theory under consideration is obtained by writing the
Lagrangian ${\cal L}^{(3|4)}$ as a  $(3|4)$-integral form integrated on the supermanifold ${\cal M}^{(3|4)}$.
As explained in \cite{Castellani:2015paa,Castellani:2014goa,Witten:2012bg} the integral form $\omega^{(3|4)}$
carries a form degree and a second quantum number known in the literature as the picture number. The latter
denotes how many delta functions of $\psi$ and of $\psi_c$ have to be included in order to integrate over the
cotangent space.
\par
To form such an integrand we use the {\it Picture Changing Operator} ${\mathbb Y}^{(0|4)}$. This a closed yet
non-exact integral form which is built in terms of differential forms of the supermanifold and in terms of
the Dirac delta functions $\delta(\psi)$ and $\delta(\psi_c)$. In the literature, the integration on
supermanifold is discussed and we do not review here, however we would like to point out that being $\psi =
d\theta$ and $\psi_c = d\theta_c$ commuting variables, they need a special prescription for the integration.
For that purpose, the integral forms are the central ingredients.
\par
We also need the differential operators ${\pmb{\iota}}_\psi$ and ${\pmb{\iota}}_{\psi_c}$, they are the
contraction operators with respect to the vector fields $D$ and $D_c$, dual to the 1-forms $\psi$ and
$\psi_c$ and they can be viewed as differential operators acting on $\psi$ and $\psi_c$. The following
general rules are valid
\begin{eqnarray}
\label{ruA} \psi \delta(\psi) =0\,,  ~~~~~~\psi_c \delta(\psi_c)=0\,, ~~~~ \psi{\pmb{\iota}}_\psi
\delta(\psi) = - \psi\,, ~~~~ \psi_c{\pmb{\iota}}_{\psi_c} \delta(\psi_c) = - \psi_c\,, ~~~~
 \end{eqnarray}
This  means that $\delta(\psi)$ and $\delta(\psi_c)$ carry no form degree, but they carry picture number (one
each delta function), but a derivative of a delta function plays the role of a negative form with positive
picture. So a generic form with maximal picture on the supermanifold $(3|4)$ considered in this paper has the
generic expression
\begin{eqnarray}
\label{ruB} \omega^{(p|4)} = f(x, \theta, \theta_c) e^{m_1} \wedge \dots \wedge e^{m_r} {\pmb{\iota}}_\psi^l
{\pmb{\iota}}_{\psi_c}^k \delta^2(\psi) \wedge\delta^2(\psi_c)
\end{eqnarray}
where $p = r - l - k$ is the total form degree and $f(x, \theta, \theta_c)$ is a generic function.
The case $p=0$ admits the cases $r=3$ and $l+k =3$, $r=2$ and $l+k=2$, $r=1$ and $l+k=1$ and
$r=l=k=0$. A simple example for a $(0|4)$-integral form reads
\begin{eqnarray}
\label{ruC}
{\mathbb Y}^{(0|4)} = \theta^2 \theta^2_c \delta^2(\psi) \wedge\delta^2(\psi_c)\,.
\end{eqnarray}
It is closed not exact. Furthermore it can also be written in a more covariant way as $\theta^4
\delta^4(\psi)$ by using 4d spinors. Its supersymmetry transformation under both  supersymmetries is
$d$-exact
\begin{eqnarray}
\label{ruCA}
\delta_\epsilon {\mathbb Y}^{(0|4)}  = d \left[ \eta^{(-1|4)}\right]\,,
\end{eqnarray}
where $\eta^{(-1|4)}$ is a $(-1|4)$-form which can be easily computed, but  its explicit form is
irrelevant for our purposes. This $(0|4)$-integral form represents the simplest example of a PCO
(in the target space, it has been introduced in pure spinor string theory in \cite{Berkovits:2004px}).
\par
The rheonomic Lagrangian described in (\ref{MM1}--\ref{MM5}), ${\cal L}^{{\cal N}=2}_{rheo}$ has the
property
\begin{eqnarray}
\label{ruD}
d {\cal L}^{(3|0)} =0\,,
\end{eqnarray}
because of the presence of auxiliary fields $D^\Lambda, \mathcal{H}^i$. Then, we can construct
the action as follows
\begin{eqnarray}
\label{ruE}
S = \int_{{\cal M}^{(3|4)}} {\cal L}^{(3|0)} \wedge {\mathbb Y}^{(0|4)}\,.
\end{eqnarray}
The integrand is a $(3|4)$-integral form, which can be integrated on the supermanifold. It
is proportional to the volume form of ${\cal M}^{(3|4)}$. Certainly, we could have looked for something
more general, such as ${\cal L}^{(3|4)}$, but in that case we would have lost the contact with the original rheonomic
Lagrangian that we want to use. A search for a more general formulation will be presented elsewhere.
\par
Being ${\mathbb Y}^{(0|4)}$ closed and not-exact, it belongs to the cohomology $H^{(0|4)}$ for which we can choose a
representative, {\it i.e.} (\ref{ruC}).  Then, choosing a different representative means
${\mathbb Y}^{(0|4)} + d \Lambda^{(-1|4)}$ and then we have
\begin{eqnarray}
\label{ruF}
S = \int_{{\cal M}^{(3|4)}} {\cal L}^{(3|0)} \wedge \left( {\mathbb Y}^{(0|4)} +  d \Lambda^{(-1|4)}\right)\,.
\end{eqnarray}
and, since ${\cal L}^{(3|0)}$ is closed, $S$ is invariant. A different representative could have additional
properties: for example a supersymmetric invariant PCO (see \cite{Grassi:2016apf} for an illustration of this
case in the context of ${\cal N}=1$ Chern-Simons theories).
\par
To begin with, if we choose the PCO given in (\ref{ruC}), we get
\begin{eqnarray}
\label{ruG}
S = \int_{{\cal M}^{(3|4)}} {\cal L}^{(3|0)} \wedge \theta^2 \theta^2_c \delta^2(\psi) \wedge\delta^2(\psi_c)\,.
\end{eqnarray}
which implies that the ${\cal L}^{(3|0)}$ is computed by setting $\theta= \theta_c = \psi= \psi_c =0$, giving
the component action (\ref{N=2stLag}) and the corresponding equations. So, inserting the easiest PCO, one
obtains the pull-back of the action on the sub-manifold ${\cal M}^{(3)} \in {\cal M}^{(3|4)}$  and all the
superfields are reduced to their first components coinciding with the physical fields.
\par
Let us now consider a different PCO of the following form:
\begin{eqnarray}
\label{ruI} {\mathbb Y}^{(0|4)} = \epsilon_{mnp} e^m\wedge e^n \left( \bar\theta \gamma^p \theta
\bar{\pmb{\iota}}_\psi \cdot{\pmb{\iota}}_\psi - \bar\theta\cdot\theta \bar{\pmb{\iota}}_\psi  \gamma^p
{\pmb{\iota}}_\psi \right) \delta^2(\psi) \delta^2(\psi_c)\,.
\end{eqnarray}
(we recall that for commuting spinors we have the following identities $\bar\psi \cdot \psi = - \bar\psi_c
\psi_c$ and $\bar\psi \gamma^p \psi = \bar\psi_c \gamma^p \psi_c$, those relations are also valid for the
contration operators ${\pmb{\iota}}_\psi$ and ${\pmb{\iota}}_{\psi_c}$, {\it i.e.} $\bar{\pmb{\iota}}_\psi
\cdot {\pmb{\iota}}_\psi = \bar{\pmb{\iota}}_{\psi_c} \cdot {\pmb{\iota}}_{\psi_c}$). To check the closure of
the PCO (\ref{ruI}), we act with the differential $d$ and we have
\begin{eqnarray}
\label{ruL} d {\mathbb Y}^{(0|4)} &=& 2 \, \epsilon_{mnp} (\bar\psi \gamma^m \psi) \wedge e^n \left(
\bar\theta \gamma^p \theta \bar{\pmb{\iota}}_\psi \cdot{\pmb{\iota}}_\psi - \bar\theta\cdot\theta
\bar{\pmb{\iota}}_\psi \gamma^p  {\pmb{\iota}}_\psi
\right) \delta^2(\psi) \delta^2(\psi_c) \nonumber \\
&+& 2 \, \epsilon_{mnp} e^m\wedge e^n \left( \bar\psi \gamma^p \theta \bar{\pmb{\iota}}_\psi
\cdot{\pmb{\iota}}_\psi - \bar\psi\cdot\theta \bar{\pmb{\iota}}_\psi  \gamma^p  {\pmb{\iota}}_\psi \right)
\delta^2(\psi) \delta^2(\psi_c)
\end{eqnarray}
If the spinors $\psi$ are free to act on Dirac delta function $\delta^2(\psi) \delta^2(\psi_c)$, the $d$
variation vanishes. On the other hand, the contraction operators ${\pmb{\iota}}_\psi$ and
${\pmb{\iota}}_{\psi_c}$ can act on them and then we have
\begin{eqnarray}
\label{ruJ}
d {\mathbb Y}^{(0|4)}
&=&
2 \, \epsilon_{mnp} {\rm Tr}(\gamma^m) \wedge e^n \left(
\bar\theta \gamma^p \theta \right) - 2 \epsilon_{mnp} {\rm Tr}(\gamma^m \gamma^p) \wedge e^n
\left(\bar\theta\cdot\theta \right) \delta^2(\psi) \delta^2(\psi_c)\nonumber \\
&+& 2\, \epsilon_{mnp} e^m\wedge e^n \left( \bar\theta\gamma^p {\pmb{\iota}}_\psi - \bar\theta \gamma^p
{\pmb{\iota}}_\psi
\right) \delta^2(\psi) \delta^2(\psi_c) \nonumber \\
&=&0
\end{eqnarray}
where the first term vanishes because of ${\rm Tr}(\gamma^m) =0$, the second term vanishes because $
\epsilon_{mnp}  {\rm Tr}(\gamma^m \gamma^p) =0$ and the third term vanishes because of the minus sign between
the two pieces. Hence the PCO presented in eq.(\ref{ruI} ) is closed and it is not exact. The presence of the
explicit $\theta$'s in such a  formula implies that it is not manifestly supersymmetric. However, its
supersymmetry variation is $d$-exact. This implies that the resulting action will not be manifestly
supersymmetric with respect to $\mathcal{N}=2$ supersymmetry, rather only with respect to $\mathcal{N}=1$
supersymmetry.
\par
To show how the superspace action is reproduced, we consider here only the kinetic terms of the matter
fields. The other pieces of the action can be derived in the same way and we do not present them here, since
they give the usual results in superspace. So, we have (here we display  only the terms that give non-trivial
contributions)
\begin{eqnarray}
\label{rupA}
&&S_{Kahler} = \int_{{\cal M}^{(3|4)}} {\cal L}^{Kahler}_{rheo}\wedge {\mathbb Y}^{(0|4)} \nonumber \\
&&= \int_{{\cal M}^{(3|4)}} \left( -2i\eta_{ij^*}\left(\overline{\chi}^{j^*}\gamma_m\chi^i\right)
\left(\overline{\psi}^c \wedge \psi^c\right)\wedge e^m
-2i\eta_{ij^*}\left(\overline{\chi}^{j^*}\chi^i\right)
\left(\overline{\psi}^c \wedge \gamma_m\psi^c\right) \wedge e^m
\right) \wedge {\mathbb Y}^{(0|4)}
\nonumber \\
&&= - 2i \int_{{\cal M}^{(3|4)}}
\left[
\eta_{ij^*}\left(\overline{\chi}^{j^*}\gamma_m\chi^i\right) (\bar \theta \gamma^m \theta) +
\eta_{ij^*}\left(\overline{\chi}^{j^*}\chi^i\right) (\bar \theta \theta) \right] e^m \wedge e^n
\wedge e^p \epsilon_{mnp} \delta^2(\psi) \delta^2(\psi_c) \nonumber \\
&&= - 2i \eta_{ij^*} \int d^3 x D^2 D^2_c
\left(\bar \theta \cdot \chi^i \overline{\chi}^{j^*} \cdot \theta \right) \nonumber \\
&& = -2i  \int d^3 [d^2\theta d^2\theta_c] \, \eta_{ij^*}  \bar z^{j^*} z^i
\end{eqnarray}
In the second line, we have picked up all the terms in the ${\cal L}^{Kalher}_{rheo}$ which are proportional
to $\bar\psi \wedge \psi \wedge e^m$ (there are several terms obtained by expanding all differentials in the
action, however, they can be collectively re-expressed in one line as above. In the third line we have used
the contraction operators ${\pmb{\iota}}_\psi$ and ${\pmb{\iota}}_{\psi_c}$ to compute the derivatives with
respect to $\psi$ and $\psi_c$. Then, we are left with a combination of the fermionic superfields $\chi^i$
and $\overline \chi^{j^*}$ and of the $\theta$'s. We have displaced all differential 1-forms $e^m$ and the
Dirac delta functions to the end of the integrand. By using a simple Fierz re-arrangment we can re-write the
action as in the forth line. There, we also made manifest the Berezin integration over $\theta$ and
$\theta_c$. In the last line we have used the identity $\bar\theta \cdot D z^i = \bar\theta \cdot \chi^i$ and
its conjugate to write the final formula. We have used the symbol $[d^2 \theta d^2\theta_c]$ to denote the
Berezin integration over $\theta$ and $\theta_c$ and the result is the correct $\mathcal{N}=2$ K\"ahler
action for the chiral matter multiplets in $\mathcal{N}=2$ superspace. The rest of the action can be derived
analogously.
\par
The present formalism encompasses all possible superspace representations of the action from the component
action to the superspace action by changing the PCO in the geometrical action (\ref{ruE}) whose constant
essential ingredient is the rheonomic action constructed according to the principles of rheonomy.
\section{Quotient singularities} We come now to the
mathematics which is of greatest interest to us, in order to address
the physical problem at stake, \textit{i.e.}, the construction of CS
theories dual to M2-branes that have the metric cone on orbifolds
$\mathbb{S}^7/\Gamma$ as transverse space. The first step is to show
that such metric cone is just $\mathbb{C}^4/\Gamma$. This is a
rather simple fact but it is of the utmost relevance since it
constitutes the very bridge between the mathematics of quotient
singularities, together with their resolutions, and the physics of
CS theories. The pivot of this bridge is the complex Hopf fibration
of the $7$-sphere.
\subsection{The complex Hopf fibration of
$\mathbb{S}^7$ and quotient singularities $\mathbb{C}^4/\Gamma$} In
order to arrive at what is for us most interesting, namely quotient
singularities of the type $\mathbb{C}^4/\Gamma$ we start from the
first of the cases listed in table \ref{sasakiani}, namely the
complex Hopf fibration of the seven sphere:
\begin{eqnarray}\label{offetto7}
    \pi & : & \mathbb{S}^7 \, \rightarrow \, \mathbb{C}\mathbb{P}^3\nonumber\\
    \forall y \in\mathbb{C}\mathbb{P}^3 & : & \pi^{-1}(y) \sim \mathbb{S}^1
\end{eqnarray}
We want to establish the following important conclusion. Writing the metric cone over the seven sphere as
$\mathbb{C}^4$, namely:
\begin{equation}\label{matrassu}
    \mathcal{C}(\mathbb{S}^7) \, =\, \mathbb{R}^8 \, \sim \, \mathbb{C}^4
\end{equation}
the homogeneous coordinates $Z^i$ of $\mathbb{C}\mathbb{P}^3$ can be identified with the standard affine
coordinates of $\mathbb{C}^4$ defined above.
\par
To this purpose we consider the standard definition of the $\mathbb{C}\mathbb{P}^3$ manifold as the set of
quadruplets $\left\{Z^1,\dots , Z^4\right\}$ modulo an overall complex factor:
\begin{equation}\label{cagnolinobasso}
    \left\{Z^1,\dots , Z^4\right\} \, \sim \, \lambda \, \left\{Z^1,\dots , Z^4\right\} \quad , \quad \forall
    \lambda \in \mathbb{C}^\star
\end{equation}
On the other hand we define the $7$-sphere as the locus in $\mathbb{C}^4$ cut out by the following
constraint:
\begin{equation}\label{paneraffermo}
    |\mathbf{Z}|^2 \, \equiv\, \sum_{i=1}^4 |Z^i|^2 \, = \, 1
\end{equation}
Let us define the K\"ahler metric on the $\mathbb{CP}^3$ in terms of the homogeneous coordinates:
\begin{equation}\label{cp3metrosco}
    ds^2_{\mathbb{CP}^3} \, = \, \frac{d\mathbf{Z}\cdot d\bar{\mathbf{Z}}}{|\mathbf{Z}|^2}\, - \,
    \frac{\left(\mathbf{Z}\cdot d\bar{\mathbf{Z}}\right)\left(\bar{\mathbf{Z}}\cdot
    d{\mathbf{Z}}\right)}{|\mathbf{Z}|^4}
\end{equation}
That the above is indeed a metric on $\mathbb{CP}^3$ is verified in the following way: if in
eq.(\ref{cp3metrosco}) $\mathbf{Z}$ is replaced by $\lambda \mathbf{Z}$  all the factors $\lambda$ and all
their differentials cancel identically. If we fix the $\lambda$-gauge by setting $Z_4\,=1$ and we rename
$Z_{1,2,3}\, = \, Y_{1,2,3}$, then we find that the above metric is identical with the K\"ahler metric
obtained from the Fubini-Study K\"ahler potential:
\begin{equation}\label{carmelitano}
    \mathcal{K}_{\mathbb{CP}^3}(\mathbf{Y}) \, = \, \log \, \left( 1+|\mathbf{Y}|^2\right)
\end{equation}
On the other hand if we consider the pull-back of the flat K\"ahler metric of $\mathbb{C}^4$ on the locus
(\ref{cagnolinobasso}) we obtain the metric of the seven sphere:
\begin{equation}\label{settenanetti}
    ds^2_{\mathbb{S}^7} \, = \, d\mathbf{Z}\cdot d\bar{\mathbf{Z}}\mid_{|\mathbf{Z}|^2\,=\,1}
\end{equation}
Let us next consider the following 1-form:
\begin{equation}\label{gattostivalus}
    \Omega \left(\mathbf{Z}\right)\, = \, \frac{\rm i}{2\,|\mathbf{Z}|^2} \left( \mathbf{Z}\cdot d\bar{\mathbf{Z}}
    \, - \,  \bar{\mathbf{Z}}\cdot d\mathbf{Z}\right)
\end{equation}
and perform the following two calculations. If we replace $\mathbf{Z}\to \lambda \mathbf{Z}$, we obtain:
\begin{equation}\label{crisappolo}
    \Omega \left(\lambda\mathbf{Z}\right)\, = \, \frac{\rm i}{2} \left (\lambda d\bar{\lambda} \, - \, \bar{\lambda}
    d\lambda\right) \, + \,\Omega \left(\mathbf{Z}\right)
\end{equation}
In particular if $\lambda \, = \, e^{\mathrm{i} \theta}$ we get:
\begin{equation}\label{connatto}
    \Omega \left(e^{\mathrm{i} \theta}\mathbf{Z}\right)\, = \, d\theta +
    \Omega \left(\mathbf{Z}\right)
\end{equation}
This shows that $\Omega$ is a $\mathrm{U(1)}$-connection on the
principal $\mathrm{U(1)}$-bundle that has $\mathbb{CP}^3$ as base
manifold and which can be identified with the $7$-sphere. The
curvature of this connection is just the K\"ahler $2$-form on
$\mathbb{CP}^3$.
\par
On the other hand we have:
\begin{equation}\label{gruttolino}
    ds^2_S \, \equiv \, d\Omega^2 + ds^2_{\mathbb{CP}^3} \, = \,\frac{d\mathbf{Z}\cdot d\bar{\mathbf{Z}}}{|\mathbf{Z}|^2}\, - \,
    \frac{\left(\mathbf{Z}\cdot d\bar{\mathbf{Z}}+\bar{\mathbf{Z}}\cdot d{\mathbf{Z}} \right)^2 }{|\mathbf{Z}|^4}
\end{equation}
If we restrict the above line element to the locus (\ref{cagnolinobasso}) we find:
\begin{equation}\label{maguardaqua}
    ds^2_S\mid_{|\mathbf{Z}|^2=1} \, = \, d\mathbf{Z}\cdot d\bar{\mathbf{Z}} \mid_{|\mathbf{Z}|^2=1}
    \, = \, ds^2_{\mathbb{S}^7}
\end{equation}
In this way we have obtained the desired result: the metric cone
over the $7$-sphere is described by the homogeneous coordinates of
$\mathbb{CP}^3$ interpreted as affine ones on $\mathbb{C}^4$:
\begin{equation}\label{gaviadinus}
    ds^2_\mathcal{C} \, = \, dr^2 \, + \, r^2 ds^2_{\mathbb{S}^7} \, = \, d\mathbf{Z}\cdot d\bar{\mathbf{Z}}
\end{equation}
Another way of stating the same result is the following one. We can regard $\mathbb{C}^4$ as the total space
of a line bundle over $\mathbb{CP}^3$:
\begin{eqnarray}\label{lignotto}
    \pi & : & \mathbb{C}^4 \, \rightarrow \, \mathbb{C}\mathbb{P}^3\nonumber\\
    \forall y \in\mathbb{C}\mathbb{P}^3 & : & \pi^{-1}(y) \sim \mathbb{C}^\star
\end{eqnarray}
The form $\Omega$ is a connection on this line-bundle.
\par
The consequence of this discussion is that if we have a finite subgroup $\Gamma \subset \mathrm{SU(4)}$,
which obviously is an isometry of $\mathbb{CP}^3$ we can consider its action both on $\mathbb{CP}^3$ and on
the seven sphere so that  we have:
\begin{equation}\label{raccimolato}
    \mathrm{AdS_4} \times \frac{\mathbb{ S}^7}{\Gamma} \, \to \partial\mathrm{AdS_4} \times
    \frac{\mathbb{C}^4}{\Gamma}
\end{equation}
We are therefore interested in describing the theory of M2-branes
probing the singularity $\frac{\mathbb{C}^4}{\Gamma}$.
\subsection{From singular orbifolds to smooth resolved manifolds} The
next point which provides an important orientation in addressing
mathematical questions  comes from physics, in view of the final use
of the considered mathematical lore in connection with M2-brane
solutions of $D=11$ supergravity and later on in the construction of
quantum gauge theories supposedly dual to such M2-solutions of
supergravity.
\par
Let us start once again from
\begin{equation}\label{caluffo2}
  K_3 \, \stackrel{\pi}{\longleftarrow} \, \mathcal{M}_7 \,  \stackrel{Cone}{ \hookrightarrow} \,  K_4 \,
  \stackrel{\mathcal{A}}{ \hookrightarrow} \, \mathbb{V}_q
\end{equation}
namely from eq. (\ref{caluffo}) that we are rewrite in slightly more
general terms. The $\mathrm{AdS_4}$ compactification of $D=11$
supergravity is obtained by utilizing as complementary
$7$-dimensional space a manifold $\mathcal{M}_7$ which occupies the
above displayed position in the inclusion--projection diagram
(\ref{caluffo2}). The metric cone $\mathcal{C}(\mathcal{M}_7)$
enters the game when, instead of looking at the vacuum:
\begin{equation}\label{vacuetto}
    \mathrm{AdS_4} \otimes \mathcal{M}_7
\end{equation}
we consider the more general M2-brane solutions of D=11 supergravity, where the D=11 metric is of the
following form:
\begin{equation}\label{m2branmet}
    ds^2_{11} \, = \, H(y)^{-\ft 23}\,\left(d\xi^\mu\otimes d\xi^\nu\eta_{\mu\nu}\right) - H(y)^{\ft 13} \,
    \left(ds^2_{\mathcal{M}_8} \right)
\end{equation}
 $\eta_{\mu\nu}$ being the constant Lorentz metric of $\mathrm{Mink}_{1,2}$ and:
\begin{equation}\label{metric8}
   ds^2_{\mathcal{M}_8} \, = \,dy^I\otimes dy^J \, g_{IJ}(y)
\end{equation}
being a Ricci-flat metric on an asymptotically locally euclidian $8$-manifold $\mathcal{M}_8$. In eq.
(\ref{m2branmet}) the symbol $H(y)$ denotes a harmonic function over the manifold $\mathcal{M}_8$, namely:
\newcommand*\DAlambert{\mathop{}\!\mathbin\Box}
\begin{equation}\label{cicio}
     \DAlambert_{g} H(y) \, = \, 0
\end{equation}
Eq.(\ref{cicio}) is the only differential constraint required in order to satisfy all the field equations of
$D=11$ supergravity in presence of  the standard M2-brain ansatz for the $3$-form field:
\begin{equation}\label{scarparotta}
    \mathbf{A}^{[3]} \, \propto \, H(y)^{-1} \left(d\xi^\mu\wedge d\xi^\nu\wedge d\xi^\rho \,
    \epsilon_{\mu\nu\rho}\right)
\end{equation}
In this more general setup the manifold $\mathcal{M}_8$ is  what substitutes the metric cone
$\mathcal{C}(\mathcal{M}_7)$. To see the connection between the two viewpoints it suffices to introduce the
radial coordinate $r(y)$ by means of the position:
\begin{equation}\label{radiatore}
    H(y) \, = \, 1 \, - \, \frac{1}{r(y)^6}
\end{equation}
The asymptotic region where $\mathcal{M}_8$ is required to be locally euclidian is defined by the condition
$r(y) \to \infty$. In this limit the metric (\ref{metric8}) should approach the flat euclidian metric of
$\mathbb{R}^8\simeq \mathbb{C}^4$. The opposite limit where $r(y)\to 0$ defines the near horizon region of
the M2-brane solution. In this region the metric (\ref{m2branmet}) approaches that of the space
(\ref{vacuetto}), the manifold $\mathcal{M}_7$ being a codimension one submanifold of $\mathcal{M}_8$ defined
by the limit $r\to 0$.
\par
To be mathematically more precise let us consider the harmonic function as a map:
\begin{equation}\label{haccusmap}
    \mathfrak{H}\quad : \quad \mathcal{M}_8 \, \rightarrow \, \mathbb{R}_+
\end{equation}
This viewpoint introduces a foliation of $\mathcal{M}_8$ into a one-parameter family of $7$-manifolds:
\begin{equation}\label{romualdo}
  \forall h \in \mathbb{R}_+  \quad : \quad \mathcal{M}_7(h) \, \equiv \,
  \mathfrak{H}^{-1}(h) \subset \mathcal{M}_8
\end{equation}
In order to have the possibility of residual supersymmetries we are interested in cases where the Ricci flat
manifold $\mathcal{M}_8$ is actually a Ricci-flat K\"ahler $4$-fold.
\par
In this way the appropriate rewriting of eq.(\ref{caluffo}-\ref{caluffo2}) is as follows:
\begin{equation}\label{caluffo3}
K_3 \, \underbrace{\stackrel{\pi}{\longleftarrow}}_{\mbox{if it applies}}  \, \mathcal{M}_7 \quad
  \stackrel{\mathfrak{H}^{-1}}{ \longleftarrow} \quad  K_4 \quad
  \stackrel{\mathcal{A}}{ \hookrightarrow} \quad \mathbb{V}_q
\end{equation}
\paragraph{The $\mathcal{N}=8$ case with no singularities.} The prototype of the above inclusion--projection
diagram is provided by the case of the M2-brane solution with all preserved supersymmetries. In this case we
have:
\begin{equation}\label{caluffopiatto}
\mathbb{CP}^3 \quad \stackrel{\pi}{\longleftarrow}  \quad \mathbb{S}^7 \quad
  \stackrel{Cone}{ \hookrightarrow} \quad  \mathbb{C}^4 \quad
  \stackrel{\mathcal{A}=\mathrm{Id}}{ \hookrightarrow} \quad \mathbb{C}^4
\end{equation}
On the left we just have the projection map of the Hopf fibration of the $7$-sphere. On the right we have the
inclusion map of the $7$ sphere in its metric cone $\mathcal{C}(\mathbb{S}^7)\equiv \mathbb{R}^8\sim
\mathbb{C}^4$. The last algebraic inclusion map is just the identity map, since the algebraic variety
$\mathbb{C}^4$ is already smooth and flat and needs no extra treatment.
\paragraph{The singular orbifold cases.} The next orbifold cases are those of interest to us in this paper and in
paper \cite{ugopietro} which will follow. Let $\Gamma \subset \mathrm{SU(4)}$ be a finite discrete subgroup
of $\mathrm{SU(4)}$. Then eq.(\ref{caluffopiatto}) is replaced by the following one:
\begin{equation}\label{calufforbo}
\frac{\mathbb{CP}^3}{\Gamma} \quad \stackrel{\pi}{\longleftarrow}  \quad \frac{\mathbb{S}^7}{\Gamma} \quad
  \stackrel{Cone}{ \hookrightarrow} \quad  \frac{\mathbb{C}^4}{\Gamma} \quad
  \stackrel{\mathcal{A}=\mbox{?}}{ \hookrightarrow} \quad \mbox{?}
\end{equation}
The consistency of the above quotient is guaranteed by the relation $\mathrm{SU(4)}\subset \mathrm{SO(8)}$.
The question marks can be removed only by separating the two cases:
\begin{description}
  \item[A)] Case: $\Gamma \subset \mathrm{SU(2)} \subset \mathrm{SU(2)_I} \otimes \mathrm{SU(2)_{II}}
  \subset \mathrm{SU(4)}$. Here we obtain:
  \begin{equation}\label{cromatorosso}
    \frac{\mathbb{C}^4}{\Gamma} \, \simeq \, \mathbb{C}^2 \otimes \frac{\mathbb{C}^2}{\Gamma}
  \end{equation}
and everything is under full control for the Kleinian
$\frac{\mathbb{C}^2}{\Gamma}$ singularities and their resolution
{\`a} la Kronheimer in terms of HyperK\"ahler quotients
\cite{kro1,kro2},\cite{ALEnostro}.
  \item[B)] Case: $\Gamma \subset \mathrm{SU(3)} \subset   \mathrm{SU(4)}$. Here we obtain:
  \begin{equation}\label{cromatonero}
    \frac{\mathbb{C}^4}{\Gamma} \, \simeq \, \mathbb{C}  \otimes \frac{\mathbb{C}^3}{\Gamma}
  \end{equation}
  and the study and resolution of the singularity $\frac{\mathbb{C}^3}{\Gamma}$ in a physics--friendly way
   is the main issue in \cite{ugopietro} which is currently on preparation.
   The comparison of case B) with the well known case A)
   is the main guide in this venture.
\end{description}
Let us begin by erasing the question marks in case A). Here we can write:
\begin{equation}\label{calufforboA}
\frac{\mathbb{CP}^3}{\Gamma} \quad \stackrel{\pi}{\longleftarrow}  \quad \frac{\mathbb{S}^7}{\Gamma} \quad
  \stackrel{Cone}{ \hookrightarrow} \quad  \mathbb{C}^2 \otimes \frac{\mathbb{C}^2}{\Gamma} \quad
  \stackrel{\mathrm{Id}\times\mathcal{A}_W}{ \hookrightarrow} \quad \mathbb{C}^2 \otimes \mathbb{C}^3
\end{equation}
In the last inclusion map on the right, $\mathrm{Id}$ denotes the identity map $\mathbb{C}^2\to \mathbb{C}^2$
while $\mathcal{A}_W$ denotes the inclusion of the orbifold $\frac{\mathbb{C}^2}{\Gamma}$ as a singular
variety in $\mathbb{C}^3$ cut out by a single polynomial constraint:
\begin{eqnarray}\label{inclusione}
  \mathcal{A}_W& : &  \frac{\mathbb{C}^2}{\Gamma}\rightarrow
  \mathbf{V}(\mathcal{I}^W_\Gamma)
  \subset \mathbb{C}^3\nonumber\\
 \mathcal{I}_\Gamma^W & \equiv & \mbox{Ideal of $\mathbb{C}[u,w,z]$ generated by
 $W_\Gamma(u,w,z)$}
\end{eqnarray}
The variables $u,w,z$ are polynomial $\Gamma$-invariant functions of
the coordinates $z_1,z_2$ on which $\Gamma$ acts linearly. The
unique generator $W_\Gamma(u,w,z)$ of the ideal
$\mathcal{I}^W_\Gamma$ which cuts out the singular variety
isomorphic to $\frac{\mathbb{C}^2}{\Gamma}$ is the unique algebraic
relation existing among such invariants. In \cite{ALEnostro}  the
relation was discussed between this algebraic equation and the
embedding in higher dimensional algebraic varieties associated with
the McKay quiver and the HyperK\"ahler quotient.
\par
Let us now consider case B). Up to this level things go in a quite
analogous way with respect to case A). Indeed we can write
\begin{equation}\label{calufforboB}
\frac{\mathbb{CP}^3}{\Gamma} \quad \stackrel{\pi}{\longleftarrow}  \quad \frac{\mathbb{S}^7}{\Gamma} \quad
  \stackrel{Cone}{ \hookrightarrow} \quad  \mathbb{C} \otimes \frac{\mathbb{C}^3}{\Gamma} \quad
  \stackrel{\mathrm{Id}\times\mathcal{A}_\mathcal{W}}{ \hookrightarrow} \quad \mathbb{C} \otimes \mathbb{C}^4
\end{equation}
In the last inclusion map on the right, $\mathrm{Id}$ denotes the identity map $\mathbb{C}\to \mathbb{C}$
while $\mathcal{A}_\mathcal{W}$ denotes the inclusion of the orbifold $\frac{\mathbb{C}^3}{\Gamma}$ as a
singular variety in $\mathbb{C}^4$ cut out by a single polynomial constraint:
\begin{eqnarray}\label{inclusioneB}
  \mathcal{A}_\mathcal{W}& : &  \frac{\mathbb{C}^3}{\Gamma}\rightarrow
  \mathbf{V}(\mathcal{I}_\Gamma^\mathcal{W})
  \subset \mathbb{C}^4\nonumber\\
 \mathcal{I}_\Gamma^\mathcal{W} & \equiv & \mbox{Ideal of
 $\mathbb{C}[u_1,u_2,u_3,u_4]$ generated by
 $\mathcal{W}_\Gamma(u_1,u_2,u_3,u_4)$}
\end{eqnarray}
Indeed as it will be shown in \cite{ugopietro}  for the cases $\Gamma\,=\, \mathrm{L_{168}}$ discussed by
Markushevich \cite{marcovaldo} and for its subgroups, the variables $u_1,u_2,u_3,u_4$ are polynomial
$\Gamma$-invariant functions of the coordinates $z_1,z_2,z_3$ on which $\Gamma$ acts linearly. The unique
generator $\mathcal{P}_\Gamma(u_1,u_2,u_3,u_4)$ of the ideal $\mathcal{I}_\Gamma$ which cuts out the singular
variety isomorphic to $\frac{\mathbb{C}^3}{\Gamma}$ is the unique algebraic relation existing among such
invariants. As for the relation of this algebraic equation with the embedding in higher dimensional algebraic
varieties associated with the McKay quiver, things are now more complicated and a thorough  discussion is
going to appear in \cite{ugopietro}.
\par
Finally let us consider the case of smooth resolutions. In case A) the smooth resolution is provided by
manifolds $ALE_\Gamma$ and we obtain the following diagram:
\begin{equation}\label{caluffoALE}
 \mathcal{M}_7 \quad
  \stackrel{\mathfrak{H}^{-1}}{ \longleftarrow} \quad  \mathbb{C}^2 \otimes ALE_\Gamma \quad
  \stackrel{\mathrm{Id}\times qK}{ \longleftarrow} \quad \mathbb{C}^2
  \otimes\mathbb{V}_{|\Gamma|+1} \quad \stackrel{\mathrm{Id}\times\mathcal{A}_{\mathcal{P}}}{\hookrightarrow}
  \quad \mathbb{C}^2 \otimes \mathbb{C}^{2|\Gamma|}
\end{equation}
In the above equation the map $\stackrel{\mathfrak{H}^{-1}}{ \longleftarrow}$ denotes the inverse of the
harmonic function map on $\mathbb{C}^2\times ALE_\Gamma$ that we have already discussed. The map
$\stackrel{\mathrm{Id}\times qK}{ \longleftarrow}$ is instead the product of the identity map $\mathrm{Id} \,
: \, \mathbb{C}^2 \to \mathbb{C}^2$ with the K\"ahler quotient map:
\begin{equation}\label{fraulein}
   qK \quad : \quad \mathbb{V}_{|\Gamma|+1} \, \longrightarrow \,\mathbb{V}_{|\Gamma|+1}\, //_{\null_K} \,
   \mathcal{F}_{|\Gamma|-1} \, \simeq \, ALE_\Gamma
\end{equation}
of an algebraic variety of complex dimension $|\Gamma|+1$ with respect to a suitable Lie group
$\mathcal{F}_{|\Gamma|-1}$ of real dimension $|\Gamma|-1$. Finally the map
$\stackrel{\mathcal{A}_{\mathcal{P}}}{\hookrightarrow}$ denotes the inclusion map of the variety
$\mathbb{V}_{|\Gamma|+1}$ in $\mathbb{C}^{2|\Gamma|}$. Let $y_1,\dots y_{2|\Gamma|}$ be the coordinates of
$\mathbb{C}^{2|\Gamma|}$. The variety $\mathbb{V}_{|\Gamma|+1}$ is defined by an ideal generated by
$|\Gamma|-1$ quadratic generators:
\begin{eqnarray}\label{congo}
    \mathbb{V}_{|\Gamma|+1} & = & \mathbf{V}\left( \mathcal{I}_\Gamma\right)\nonumber\\
 \mathcal{I}_\Gamma &= & \mbox{Ideal of $\mathbb{C}\left[y_1,\dots y_{2|\Gamma|}
\right]$ generated by
$\left\{\mathcal{P}_1(y),\mathcal{P}_2(y),\dots
,\mathcal{P}_{|\Gamma|-1}(y)\right\}$}
\end{eqnarray}
Actually the $|\Gamma|-1$ polynomials $\mathcal{P}_i(y)$ are the holomorphic part of the triholomorphic
moment maps associated with the triholomorphic action of the group $\mathcal{F}_{|\Gamma|-1}$ on
$\mathbb{C}^{2|\Gamma| }$ and the entire procedure from  $\mathbb{C}^{2|\Gamma|}$ to $ALE_\Gamma$ can be seen
as the HyperK\"ahler quotient:
\begin{equation}\label{cagliozzo}
    ALE_\Gamma \, = \,\mathbb{C}^{2|\Gamma|}//_{\null_{HK}}\mathcal{F}_{|\Gamma|-1}
\end{equation}
yet we have preferred to split the procedure into two steps in order to compare case A) with case B) where
the two steps are necessarily distinct and separated.
\par
Indeed in case B) we can write the following diagram:
\begin{equation}\label{caluffoBalle}
 \mathcal{M}_7 \quad
  \stackrel{\mathfrak{H}^{-1}}{ \longleftarrow} \quad  \mathbb{C}^1 \otimes Y_\Gamma \quad
  \stackrel{\mathrm{Id}\times qK}{ \longleftarrow} \quad \mathbb{C}
  \otimes\mathbb{V}_{|\Gamma|+2} \quad \stackrel{\mathrm{Id} \times\mathcal{A}_{\mathcal{P}}}{\hookrightarrow}
  \quad \mathbb{C} \otimes\mathbb{C}^{3|\Gamma|}
\end{equation}
In this case, just as in the previous one,  the intermediate step is provided by the K\"ahler quotient but
the map on the extreme write $\stackrel{\mathcal{A}_{\mathcal{P}}}{\hookrightarrow}$ denotes the inclusion
map of the variety $\mathbb{V}_{|\Gamma|+2}$ in $\mathbb{C}^{3|\Gamma|}$. In this case the definition of the
variety $\mathbb{V}_{|\Gamma|+2}$ is a more complicated issue and it will be presented in \cite{ugopietro}.
\section{Conclusions}
In the present paper we have considered the general form of D=3 $\mathcal{N}=2$ gauge theories from three
point of views:
\begin{enumerate}
  \item The \textit{structural point of view}, meaning with this the application of the various approaches to the construction of
  supersymmetric field theories and their relation. In this context we have illustrated the use of the
  \textit{integral form formalism} and the extraordinary conceptual advances encoded in the notion of the \textit{Picture
  Changing Operators}. It appears in general and it was effectively illustrated in the present case that the
  time-honored \textit{rheonomic lagrangian} includes in an implicit way all the other formalisms, the
  component formalism, the various superfield formalisms and so on. One goes from one formalism to other
  changing representatives of cohomology classes  within a new sophisticated setup of cohomological-algebras
  associated with supermanifolds that was developed in
  \cite{Castellani:2015paa,Catenacci:2016qzd,Grassi:2016apf,Castellani:2016ibp}. Establishing  algorithmic transitions back and
  forth from the
  component-like approach, which is better suited to geometrical visions, to the superfield approach, which is
  better suited for quantum calculations, is a new added value, possibly quite relevant in relation with the
  next aspects of the supersymmetric Chern-Simons theories here discussed.
  \item The \textit{geometrical point of view} meaning with this the upgrading of the \textit{rheonomic
  lagrangian} of D=3 matter coupled gauge theories originally constructed in \cite{Fabbri:1999mk}. That
  lagrangian has now been rewritten in more general terms utilizing an arbitrary K\"ahler potential $\mathcal{K}(z,{\bar z})$
  for the Wess-Zumino multiplet, with an arbitrary Lie group of holomorphic isometries and an arbitrary
  superpotential $W(z)$. This fully general off-shell construction in the rheonomic approach was so far
  lacking and the present paper fills the gap. In view of what stated in point i) this is particularly
  important.
  \item The point of view of applications to the \textit{$\mathrm{AdS_4/CFT_3}$} correspondence and the
  involved \textit{algebro--geometric issues}. In the context of the general type of theories described above we have
  reconsidered the issue of the construction of three-dimensional theories dual to $D=11$ supergravity compactified
  on $\mathrm{AdS_4}\times \mathcal{M}_7$  or, better said, to $M2$-brane solutions admitting a K\"ahlerian non
  compact four-fold $K_4$ as transverse space to the brane world--volume. Recalling that the particular type
  of  Chern-Simons gauge theories used in the ABJM-like models arises from the gaussian integration of
  auxiliary fields and of those physical scalars that at the infrared fixed point loose their kinetic terms,
  according to a mechanism which was discovered much earlier than ABJM in \cite{Fabbri:1999ay,Fre':1999xp,ringoni}
  and there fully utilized, we focused on the ample class of cases where
  $K_4$ is obtained as a K\"ahler or HyperK\"ahler quotient from a larger flat (Hyper)K\"ahler variety $\mathbb{V}_q$.
  Using as examples the classified 7-dimensional sasakian coset manifolds $\mathrm{G/H}^{sasak}_7$, we have stressed
  that the data for the construction of the searched for world-volume gauge theory are essentially the same as the
  geometrical data of the K\"ahler quotient construction of the transverse space.
  \par
  Apart from the case of  metric cones $\mathcal{C}\left(\mathrm{G/H}^{sasak}_7\right)$ over sasakian
  homogeneous spaces another relevant case of K\"ahler quotients is provided by the crepant resolutions of
  singular $\mathbb{C}^n/\Gamma$ orbifolds where $\Gamma\subset \mathrm{SU(n)}$ is a finite subgroup of the
  relevant unitary group. In this case a generalization of the Kronheimer construction of ALE manifolds and a
  generalization of the McKay correspondence play a crucial role in the resolution of the singularity via
  K\"ahler quotient and consequently in the construction of the $D=3$ Chern Simons gauge theory. The subtle
  and exciting mathematical aspects of these construction constitute the target of the forthcoming paper
  \cite{ugopietro}: in the present paper we have anticipated the general scheme how the geometrical data
  of  singularity resolutions are to be utilized within the context of D=3 gauge theory constructions.
\end{enumerate}

\section*{Acknowledgements}
One of us (P.F.) acknowledges important clarifying discussions with his good friend and coauthor Ugo Bruzzo
and with his long time collaborator Aleksander Sorin, particularly at the beginning of the present project.
The two of us are grateful to our frequent collaborators and excellent friends Mario Trigiante, Leonardo
Castellani, Carlo Maccaferri and Roberto Catenacci for innumerable always precious conversations and discussions.

\end{document}